\documentclass[prb,showpacs,amsmath,amssymb,twocolumn]{revtex4}

\usepackage[pdftex]{graphicx}
\usepackage[pdftex]{epsfig}
\usepackage{epstopdf}
\usepackage{dcolumn}
\usepackage{bm}
\usepackage{color}

\graphicspath{{converted_graphics/}}
\begin{document}

\title{Effective mass theory of monolayer $\delta$-doping in the high-density limit}

\author{Daniel W. Drumm}
\email{d.drumm@pgrad.unimelb.edu.au}
\affiliation{Centre for Quantum Computation and Communication Technology, School of Physics, The University of Melbourne, Parkville 3010, Australia}
\author{Lloyd C. L. Hollenberg}
\affiliation{Centre for Quantum Computation and Communication Technology, School of Physics, The University of Melbourne, Parkville 3010, Australia}
\author{Michelle Y. Simmons}
\affiliation{Centre for Quantum Computation and Communication Technology, University of New South Wales, Sydney NSW 2052, Australia}
\author{Mark Friesen}
\affiliation{Department of Physics, University of Wisconsin-Madison, Wisconsin 53706, USA}

\begin{abstract}
Monolayer $\delta$-doped structures in silicon have attracted renewed interest with their recent incorporation into atomic-scale device fabrication strategies as source and drain electrodes and in-plane gates.  Modeling the physics of $\delta$-doping at this scale proves challenging, however, due to the large computational overhead associated with \textit{ab initio} and atomistic methods.  Here, we develop an analytical theory based on an effective mass approximation.  We specifically consider the Si:P materials system, and the limit of high donor density, which has been the subject of recent experiments.  In this case, metallic behavior including screening tends to smooth out the local disorder potential associated with random dopant placement.  While smooth potentials may be difficult to incorporate into microscopic, single-electron analyses, the problem is easily treated in the effective mass theory by means of a jellium approximation for the ionic charge.  We then go beyond the analytic model, incorporating exchange and correlation effects within a simple numerical model.  We argue that such an approach is appropriate for describing realistic, high-density, highly disordered devices, providing results comparable to density functional theory, but with greater intuitive appeal, and lower computational effort.  We investigate valley coupling in these structures, finding that valley splitting in the low-lying $\Gamma$ band grows much more quickly than the $\Gamma$-$\Delta$ band splitting at high densities.  We also find that many-body exchange and correlation corrections affect the valley splitting more strongly than they affect the band splitting.
\end{abstract}


\maketitle

\section{Introduction}
As transistors continue to shrink in size, they approach a limiting regime where the electrons are confined and controlled over atomic length scales.  Silicon-based devices have shown particular promise in this regard.  For example, electrons on individual donors or traps have been probed inside conventional metal on insulator field-effect transistors,\cite{Xiao2004,Morello2010,Lansbergen2008,Huebl2010} and have been proposed as qubits for quantum computing architectures. \cite{Kane98,Vrijen00,Skinner03,Friesen03,Schofield03,Hollenberg04,Hill05}  In other experiments, quantum dots with fewer than ten deterministically-positioned donors have been tunnel-coupled to proximal leads,\cite{martin} and single electrons have been confined using electrostatic top-gates.\cite{Simmons2007}  In several experiments, individual spins have also been measured.\cite{Xiao2004,Morello2010,Simmons2011}  While such technologies will have applications for conventional computing, much of the recent progress in this area has been spurred by the quest for spin-based qubits.\cite{Loss98,Kane98,Elzerman2004,Petta2005,Shaji08}

In this work, we focus on devices formed of degenerately doped phosphorus in silicon.  In the laboratory, the silicon is masked by hydrogen atoms, which are lithographically patterned using a scanning tunneling microscope.\cite{Wada1997,Tucker1998,Ruess2004,Shen2004}  Phosphorus atoms from phosphine gas are then incorporated into unmasked segments of the top (monatomic) layer of silicon.  A self-limiting growth mechanism leads to rather uniform doping densities corresponding to one substitutional donor for every four atomic sites in the $\delta$-layer.\cite{Wilson04}  The resulting devices are fully epitaxial.

Realistic theoretical models of Si:P disorder have proven challenging, with calculated band structures known to depend very sensitively on the disorder model.
In density functional calculations, for example, the number, the splitting, or even the existence of energy bands depends on the nature of the symmetries, the placement of the donors and the size of the computational unit cell.\cite{Carter2011}  In addition to disorder, fabrication geometries play an important role in device operation.  Specialized structures in typical devices may range in size from nanometers to microns,\cite{martin} causing technical challenges for any theoretical treatment.  It is not currently feasible to treat large, disordered devices with atomistic accuracy; hybrid approaches, however, may provide viable, self-contained solutions for such large-scale problems.  Multi-scale methods, in particular, hold great promise.  Examples include the merging of tight binding and local density techniques.\cite{Hoon2009}  There are also computational advantages to eliminating disorder effects within the doping plane.  In this case, translational symmetry can be restored by averaging.\cite{qian05,Carter2011}

In this paper, we develop a coarse-grained theory of $\delta$-doped Si:P devices, consistent with the effective mass approximation.  Effective mass theory (EMT) provides an efficient means for analyzing large, complex systems, like tunnel-coupled devices.  Recently, such methods were applied to the problem of few-electron quantum dots.\cite{martin}  The main modification to the bulk EMT, required for $\delta$-doping, involves the uniform shifting of energy bands, up or down.  These shifts account for the quantum confinement in the $\delta$-doping potential.  A closely related system, whose band structure can be understood in terms of band shifting, is the inversion layer.\cite{AFS}  An important distinction between inversion layer and Si:P devices is that the latter have very high carrier densities, which leads to the occupation of multiple bands.  Density functional methods confirm this picture of simple energy shifts in disordered Si:P geometries.\cite{qian05,oliver}  It is especially important to note that the parabolic shape of the bands and the corresponding curvature (the so-called inverse effective mass) are largely unaffected by the energy shifts.  This suggests a straight-forward modification of the bulk effective mass theory to incorporate the confinement effects, which we describe in detail.

We consider several approaches for obtaining a two-dimensional (2D)-EMT in Si:P.  The simplest approach is empirical.  In this case, the relevant parameters in the 2D-EMT, which we can think of as inputs to the theory, are obtained from a more rigorous, \emph{ab initio} band structure calculation.  When possible, the parameters are  obtained directly from experiments.  We go on to describe the physical features and phenomena that may be computed within the EMT.  These include many-body effects, exchange and correlation, and valley splitting.  The question of disorder and donor placement is not a concern for the EMT because of the coarse-grained nature of the theory.  The characteristic length scale associated with the coarse-graining is the effective Bohr radius.  For quarter-monolayer doping, this length scale encompasses many donors.  A jellium approximation for describing the donor charge is therefore appropriate to our problem.  Local variations of the jellium density could be introduced into this theoretical framework; we do not, however, consider such problems here.

The theoretical approach we employ is similar to the numerical effective mass theory of Scolfaro, \emph{et al.}\cite{Scolfaro94} who consider a periodic array of thick low-density $\delta$-layers. The larger separation between their donors required a 3D treatment and produced valley splittings of the $\Delta$-band in the range of 20-46 meV, which is more consistent with individual donors than high-density $\delta$-layers, where the $\Delta$-band is essentially degenerate.\cite{Carter2011}  These increased with doping density and are inconsistent with \emph{ab initio} results for high-density $\delta$-layers.\cite{Carter2011}  Here, we treat the experimentally relevant case of thin layers of high doping density, where it is possible to project the 3D-EMT onto a 2D theory of immediate interest for 2D devices.  As much as possible, we focus on analytical (rather than numerical) methods, which allows us to identify the underlying physics of the $\delta$-layers.  For example, we obtain a density scaling theory.  Later, we obtain numerical solutions that provide theoretical parameters for the 2D theory.  The present analysis is formulated in terms a single $\delta$-layer; although our geometry is not periodic, multiple layers could be treated by a straightforward extension of our theory.  Rodriguez-Vargas \emph{et al.} \cite{Rodriguez-Vargas06} also solved a non-periodic double-layer system numerically, although their approach is semi-classical where ours is fully quantum mechanical.

The paper is outlined as follows.  In Sec.~\ref{sec:2DEMT} we identify input parameters and develop a description for the shifting and filling of the bands within the EMT.  In Sec.~\ref{sec:formalism}, we clarify the main concepts of the shifted band model by deriving the $\delta$-doping EMT from a bulk, 3D-EMT.  This provides a setting to discuss the various components of the theory, which are not normally associated with EMT, but may be easily incorporated.  These include many-body interactions, valley splitting, exchange and correlation effects.  The utility of the effective mass method is demonstrated in Sec.~\ref{sec:variational}, where an analytical, variational theory of $\delta$-doping is derived.  This leads naturally to a scaling theory for the quantum confinement lengths and the energies of the different conduction bands.  We also perform a more rigorous, numerical analysis of the $\delta$-doping problem, to obtain an alternative set of EMT parameters, which we compare to results from more microscopic derivations in Sec.~\ref{sec:methods}.  We conclude in Sec.~\ref{sec:conc}.  The two appendices provide further details on (A) the numerical methods employed in our work, and (B) our exchange-correlation analysis.

\section{Two-Dimensional Effective Mass Theory} \label{sec:2DEMT}

\begin{figure}[t] 
  \centering
  \includegraphics[width=3.1in,keepaspectratio]{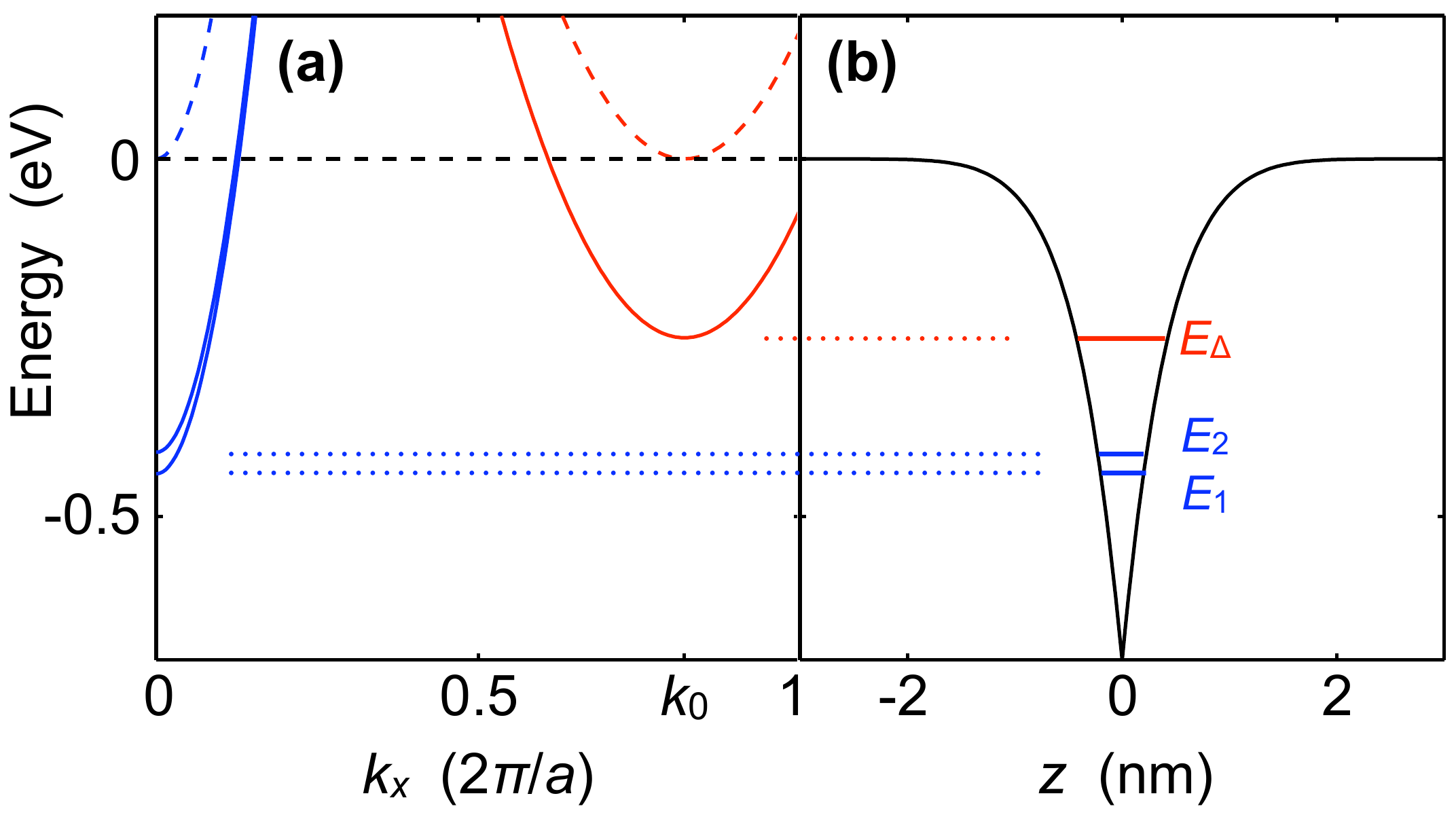}
  \caption{(Color online) 
  (a) Effective mass theory for bulk Si (dashed lines) and for $\delta$-doped Si:P (solid lines).  The minima of the bulk conduction bands define the energy zero.  (b) The donors in the $z=0$ plane produce a laterally-averaged electrostatic confinement potential, with the predominant eigenstates $\Gamma_1$, $\Gamma_2$, and $\Delta$.  The confinement along $z$ leads to transverse bands along $k_x$ and $k_y$ that are shifted downward into the gap, as shown in (a).  For all the plots, we take $k_y=0$.  For the bulk $\Gamma$ bands (dashed blue curve) we also take $k_z=k_0$, while for the bulk $\Delta$ bands (dashed red curve) we take $k_z=0$.  The various bands are then projected onto the same axis.  For the $\delta$-doped bands, the energy shifts are determined by the eigenstates shown in (b).}
  \label{fig:Energies}
\end{figure}

In this section, we describe the modifications to the bulk conduction band structure of Si, due to $\delta$-doping in the $z=0$ plane.  For $n$-type devices in the low temperature regime, the active electrons tend to fill only the low energy portions of the conduction band, known as valleys.  As is well known,\cite{DaviesBook} the valleys in bulk silicon are six-fold degenerate, with minima occurring in the equivalent $[100]$ directions, about 85\% of the way to the Brillouin zone boundary.  A given valley minimum therefore occurs at $k_0\simeq 0.85 (2\pi /a)$, where $a=0.543$~nm is the length of the cubic unit cell.  To a good approximation, the low energy band structure in a given valley appears parabolic.  For example, for the $+x$ valley (along $[100]$),
\begin{equation}
E_{+x}\simeq \frac{\hbar^2(k_x-k_0)^2}{2m_l} +
\frac{\hbar^2k_y^2}{2m_t}+\frac{\hbar^2k_z^2}{2m_t} +E_c, \label{eq:Epx}
\end{equation}
where $m_l\simeq 0.92m_0$ is the longitudinal effective mass, $m_t\simeq 0.19m_0$ is the transverse effective mass, $m_0$ is the bare electron mass, and $E_c$ is the conduction band minimum.  A cut through $E_{+x}({\bm k})$ along $[100]$ is sketched as a dashed, red curve in Fig.~\ref{fig:Energies}(a).  Here, we have set $E_c=0$, defining the band minimum as the zero of the energy.  Although the band structure associated with the $y$ and $z$ valleys lies outside the range of the plot in Fig.~\ref{fig:Energies}(a), their wavevectors also have $k_x$ components, which can be projected onto the $k_x$ axis as shown with a dashed, blue line.  These projected valleys are centered at $k_x=0$, analogous to the transverse terms in Eq.~(\ref{eq:Epx}).  The anisotropic effective masses, $m_l$ and $m_t$, together with the valley minima at $k_0$, capture the main low-energy physics of this problem, and they form the main inputs to the 3D bulk effective mass theory.

A single $n$-type donor ion, such as P, creates a local dip in the electrostatic potential, with a corresponding low-energy bound state 46~meV below the conduction band.  For $\delta$-doping in the $z=0$ plane, an electronic wave function extends over many donors.  The electrostatic potential and corresponding binding energies are much deeper than for a single donor.  Because the electron covers many randomly placed donors, it is convenient to ignore their individual positions, and to instead treat the dopants through a 2D ``jellium'' approximation, corresponding to a totally uniform charge distribution in the $z=0$ plane equal to the average 2D charge density of the discrete dopants.  For an infinite sheet of charge, the electrostatic potential does not vary in the lateral plane.  Figure~\ref{fig:Energies}(b) shows a vertical cut $V(z)$ through such an electrostatic potential, together with the resulting confined states.  Assuming overall charge neutrality, $V(z)$ flattens out far away from the doping plane.  The correct binding energy is obtained by aligning this asymptotic potential with the bottom of the bulk conduction band, as shown in the figure.

The result of the vertical confinement $V(z)$ is to reduce the continuum of bulk Bloch states from 3D to 2D.  Specifically, the allowable $k_z$ components of the Bloch states become quantized to form bound states.  The parabolic energy decomposition of Eq.~(\ref{eq:Epx}) suggests that the third term on the right-hand-side (the $z$ term) should be replaced by a quantized band energy:
\begin{equation}
E_{+x}\simeq \frac{\hbar^2(k_x-k_0)^2}{2m_l} +
\frac{\hbar^2k_y^2}{2m_t}+E_\Delta. \label{eq:Epx2D}
\end{equation}
Similarly for the $+y$ and $+z$ valleys, we have
\begin{eqnarray}
E_{+y} &\simeq& \frac{\hbar^2 k_x^2}{2m_t} +
\frac{\hbar^2(k_y-k_0)^2}{2m_l}+E_\Delta , \label{eq:Epy2D} \\
E_{+z} &\simeq& \frac{\hbar^2(k_x^2+k_y^2)}{2m_t} +E_\Gamma . \label{eq:Epz2D}
\end{eqnarray}
Analogous equations are obtained for the $-x$, $-y$, and $-z$ valleys by replacing $k_0\rightarrow -k_0$.  The quantized energies $E_\Gamma$ and $E_\Delta$ depend on the effective masses in the $k_z$ terms of the bulk equations.  The $x$-$y$ valleys are 8-fold degenerate, including spin and valley degeneracies, while the $z$ valleys are 4-fold degenerate.  Later, we will discuss the lifting of the $\Gamma$ degeneracy, in a process known as valley splitting.  (For most cases of interest, the lifting of the $\Delta$ degeneracy, if present, will be negligible.)  This leads to the distinct quantized energies, $E_1$ and $E_2$, corresponding to the $\Gamma_1$ and $\Gamma_2$ bands, which are shown in Fig.~\ref{fig:Energies}(b).  The leading order effect of vertical confinement is therefore to reduce the 3D band structure to 2D, as shown in panel (a), with the 2D bands shifted downward by their respective binding energies.

In the arguments presented above, the effective masses in the 2D theory should be identical to the bulk effective masses.  Any deviations from the energy decomposition of Eq.~(\ref{eq:Epx}) would imply mixing the effective masses and weaken the theory.  Rigorous \emph{ab initio} band structure calculations with a laterally averaged charge distribution confirm that the 2D effective masses for Si:P with the most prevalent 1/4~monolayer (ML) doping level are almost identical to the bulk masses in Si,\cite{qian05} and that the bulk bands are simply shifted downward by their respective binding energies. \cite{oliver}  In short, all the evidence suggests that 2D and 3D effective mass theories should both be accurate in this system.

The main parameters characterizing the 2D effective mass theory are $E_1$, $E_2$, $E_\Delta$, $m_t$, $m_l$, and $k_0$.  Preferably, their values should be obtained from experiments, or if not, from accurate \emph{ab initio} theories.  In a later section, we will show that $E_1$, $E_2$, and $E_\Delta$ can be derived directly from the 3D effective mass theory.  Additional derived quantities of interest include the fractional fillings of the different conduction bands.  These fillings have important implications for processes like transport and tunneling, which may occur much more readily in one band than another.  While the total filling of the 2DEG is determined by the total number of electrons, or the ionic charge density (assuming charge neutrality), the problem of calculating fractional band fillings is much subtler, since it depends on the accuracy of the binding energy calculations.

The respective filling fractions are defined as $\beta_1$, $\beta_2$ and $\beta_\Delta$, where
\begin{equation}
\beta_1+\beta_2+\beta_\Delta =1 . \label{eq:betasum2}
\end{equation}
Alternatively, combining the $\Gamma_1$ and $\Gamma_2$ bands into a single $\Gamma$ band gives
\begin{equation}
\beta_\Gamma+\beta_\Delta =1 . \label{eq:betasum}
\end{equation}
The corresponding charge densities are given by $\sigma_\gamma =-\beta_\gamma \sigma $, where $\gamma=1$, 2 or $\Delta$, and $\sigma$ is the average ionic charge density.  For 1/4~ML doping, we have $\sigma=0.2717$~C/m$^2$.  Conventional 2D band filling arguments lead to
\begin{gather}
E_F-E_1 = \beta_1 \frac{\pi \sigma \hbar^2}{em_t} ,\label{eq:D1} \\
E_F-E_2  = \beta_2 \frac{\pi \sigma \hbar^2}{em_t} ,\label{eq:D2}  \\
E_F-E_\Delta  = \beta_\Delta \frac{\pi \sigma \hbar^2}{4e\sqrt{m_tm_l}} , \label{eq:DD} 
\end{gather}
where we assume a shared chemical potential $E_F$ at zero temperature.  (Non-zero temperatures could also be considered, although very low temperatures are assumed here, as appropriate for the main applications of interest.)  For a combined $\Gamma$ band we have
\begin{equation}
E_F-E_\Gamma = \beta_\Gamma \frac{\pi \sigma \hbar^2}{2em_t} .\label{eq:DG} 
\end{equation} 
The linear system of equations (\ref{eq:betasum2}), and (\ref{eq:D1})-(\ref{eq:DD}) [or (\ref{eq:betasum}), (\ref{eq:DD}), and (\ref{eq:DG})] may readily be solved to obtain the filling fractions and $E_F$.  For example, the $\Gamma$-$\Delta$ solution is given by
\begin{gather}
E_F=\frac{ E_\Delta + E_\Gamma\sqrt{{m_t}/{4m_l}}
+ ({\pi \sigma \hbar^2}/{4e\sqrt{m_tm_l}})}
{1+\sqrt{{m_t}/{4m_l}}} , \label{eq:muF} \\
\beta_\Gamma=\frac{ E_\Delta - E_\Gamma
+ ({\pi \sigma \hbar^2}/{4e\sqrt{m_tm_l}})}
{\left( 1+\sqrt{{m_t}/{4m_l}}\right) ({\pi \sigma \hbar^2}/{2em_t})} , \label{eq:be}
\end{gather}
with $\beta_\Delta=1-\beta_\Gamma$.

To conclude this section, we note that the 2D effective mass theory, described above, includes only the six low-lying valleys of the bulk conduction band structure.  It is known that other bands may also begin to fill at the 1/4~ML doping level; most notably, the $1X$/$2X$ band may dip slightly below the Fermi energy.\cite{qian05}  These bands could be included in our 2D theory using the same methods described above.  Their fractional fillings, however, are small enough that the main physics is already captured in the theory as presented here.
%
%

\section{Three-Dimensional Treatment of $\bm \delta$-Doping}
\label{sec:formalism}
In this section, we use the 3D-EMT to study the problem of $\delta$-doping in the $z=0$ plane.  Due to our assumption of uniform doping in the lateral plane, the calculation is one-dimensional in the variable $z$.  Our ultimate goal is to derive the input parameters for a 2D-EMT.  

The theory must include many-body effects, and we begin by developing a simple Hartree theory.  We go on to obtain a variational solution to the problem of $\delta$-doping in Si, as well as scaling estimates for the vertical confinement lengths and energies in the $\Gamma$ and $\Delta$ bands.  We show how valley splitting can readily be included as a correction to the effective mass theory.  Finally, we extend the theory to include exchange and correlation contributions, which are used to obtain more accurate results for the 2D-EMT.

\subsection{Hartree Theory}

Within the jellium approximation, the ionic charge density is given by
\begin{equation}
\rho_i (z) = \sigma \delta (z) , \label{eq:rhoi}
\end{equation}
where we take $z=0$ as the $\delta$-doping plane.  In an effective mass-Hartree theory,\cite{DaviesBook} the electron charge densities in the $\Gamma$ and $\Delta$ bands are defined as
\begin{eqnarray}
\rho_\Gamma (z) &=& -\sigma \beta_\Gamma F_\Gamma^2(z) , \label{eq:rhoG} \\
\rho_\Delta (z) &=& -\sigma \beta_\Delta F_\Delta^2(z) , \label{eq:rhoD}
\end{eqnarray}
where $F_\Gamma(z)$ and $F_\Delta(z)$ are the corresponding envelope functions obtained by solving  the Schr\"{o}dinger-like equations 
 \begin{gather}
\left[ -\frac{\hbar^2}{2m_l}\frac{d^2}{dz^2}+V(z) \right] F_\Gamma (z) 
= \varepsilon_\Gamma F_\Gamma (z)  ,  \label{eq:epG} \\
\left[ -\frac{\hbar^2}{2m_t}\frac{d^2}{dz^2}+V(z) \right] F_\Delta (z) 
= \varepsilon_\Delta F_\Delta (z) .
\label{eq:epD} 
\end{gather}
Note that the full electronic wavefunctions also involve fast oscillations (\emph{e.g.}, Bloch oscillations) that occur over atomic length scales.  These fast oscillations do not enter the envelope function equations (\ref{eq:epG}) and (\ref{eq:epD}).\cite{Kohn}  In Sec.~\ref{sec:VS}, below, we investigate perturbative corrections to the energy that occur when fast oscillations are taken into account in the $\Gamma$ bands.  At the present level of approximation, however, $\Gamma_1$ and $\Gamma_2$ have the same effective mass and the same envelopes.  For now, we therefore consider just a single $\Gamma$ band.  Equations~(\ref{eq:betasum}) and (\ref{eq:rhoi})-(\ref{eq:rhoD}) explicitly satisfy charge neutrality when the envelope functions are properly normalized.  The full 3D charge density is then given by $\rho(z) =\rho_i(z) +\rho_\Gamma(z) +\rho_\Delta(z)$; by design, we obtain $\int\! \rho(z)=0$.

The electrostatic potential is calculated from Poisson's equation and Eqs.~(\ref{eq:rhoi})-(\ref{eq:rhoD}).  It is convenient to compute the electric field contributions from each of the different charge sources.  Making use of charge uniformity in the lateral plane, we obtain
\begin{eqnarray}
{\bf E}_i (z) &=& \frac{ \hat{\bf z} \sigma}{2\epsilon} \text{sign}(z) , \label{eq:Ei} \\
{\bf E}_\gamma (z) &=& \frac{\hat{\bf z}}{\epsilon} \int_0^z \rho_\gamma \, dz  , \label{eq:Egam}
\end{eqnarray}
where 
we have adopted the convention that $\int_0^\infty \delta(z)\, dz=1/2$.  Note that we will adopt the dielectric value $\epsilon=11.4\,\epsilon_0$ throughout this work, as appropriate for silicon at very low temperatures.

The corresponding contributions to the electrostatic confinement potential are then given by
\begin{gather}
V_i (z)= \frac{e\sigma |z|}{2\epsilon} ,  \label{eq:Vi}  \\
V_\gamma (z) =e\int_0^z E_{\gamma z} (z)\, dz , \label{eq:Vgam} 
\end{gather}
where $E_{\gamma z} = {\mathbf{E}}\cdot {\mathbf{z}}$.
The total Hartree potential, used in Eqs.~(\ref{eq:epG}) and (\ref{eq:epD}), is then given by $V(z)=V_i(z)+V_\Gamma(z)+V_\Delta(z)$.  Note that we have adopted the energy normalization $V(0)=V_i(0)=0$.  In this way, the potential minimum remains anchored, and is not affected by the specific details of the electronic wavefunctions.  Such considerations simplify our variational calculation in Sec.~\ref{sec:variational}, and are analogous to the Fang-Howard procedure used for inversion layers.\cite{DaviesBook}  (Later, for presentation purposes, we will renormalize the energy as in Fig. \ref{fig:Energies} such that the electrostatic potential is aligned with the bulk conduction band, in the region far from the $\delta$-doped layer.)  In principle, the methods described here could be modified to include external gates -- for example, by introducing an external electric field.  We do not consider this possibility here.

In the Hartree many-body method, we must obtain self-consistent solutions for the envelope functions (\ref{eq:epG}) and (\ref{eq:epD}), the Hartree potentials (\ref{eq:Vi}) and (\ref{eq:Vgam}), and the Fermi level (\ref{eq:muF}).  The electron energies $E_\Gamma$ and $E_\Delta$ that appear in Eqs.~(\ref{eq:muF}) and (\ref{eq:be}), however, correspond to band minima; they are not the same as the single-particle energies $\varepsilon_\Gamma$ and $\varepsilon_\Delta$ that appear in Eqs.~(\ref{eq:epG}) and (\ref{eq:epD}).  The band minima are given by
\begin{eqnarray}
E_\Gamma &=& \langle {T}[m_l] \rangle_\Gamma +\langle V_i\rangle_\Gamma+\frac{1}{2}\langle V_\Gamma \rangle_\Gamma +\langle V_\Delta \rangle_\Gamma , \label{eq:EG} \\
E_\Delta &=& \langle {T}[m_t] \rangle_\Delta +\langle V_i\rangle_\Delta +\langle V_\Gamma \rangle_\Delta
+\frac{1}{2}\langle V_\Delta \rangle_\Delta , \label{eq:ED}
\end{eqnarray}
where ${T}[m^*]$ are the same kinetic energy operators appearing in Eqs.~(\ref{eq:epG}) and (\ref{eq:epD}), and the subscripts $\Gamma$ and $\Delta$ refer to single-particle wavefunctions used to compute the expectation values.  The prefactors of $1/2$ in the Hartree terms prevent over-counting of the electron-electron interactions. \cite{DaviesBook}  We emphasize that Eqs.~(\ref{eq:EG}) and (\ref{eq:ED}) describe the band minima, and do not include any lateral kinetic energy.  The lateral kinetic energy appears explicitly in the Fermi level equations.

\subsection{Variational Calculation}
\label{sec:variational}

One of the main benefits of an effective mass theory is its simplicity and the ease with which solutions can be obtained.  We take advantage of this now to obtain initial estimates for the band minima and the electron wavefunctions, by means of a variational method.  We may even obtain simple analytical estimates, which allow us to scale the electron eigenfunctions and energy values.

The problem can be formulated in several ways.  Here, we consider a simple variational form for the single-electron wavefunctions, which is generally found to be consistent with more accurate treatments:
\begin{equation}
F_\gamma(z) = \left( \frac{2}{\pi a_\gamma^2} \right)^{1/4} e^{-(z/a_\gamma)^2} . \label{eq:FGauss}
\end{equation}
The wavefunction widths $a_\Gamma$ and $a_\Delta$, and the filling fractions $\beta_\Gamma$ and $\beta_\Delta$, represent the variational parameters in this approach,
although we will use Eq.~(\ref{eq:betasum}) to eliminate one of these variables ($\beta_\Delta$).  The Gaussian form is particularly effective in such a variation calculation because of its simplicity, and because it captures the essential properties of the wavefunction (the width), while the maximum value of the wavefunction is correctly determined via normalization.  The Gaussian tail decays too quickly compared with a more realistic wavefunction; the tail, however, contributes very little to the expectation values in Eqs.~(\ref{eq:EG}) and (\ref{eq:ED}), and therefore does not affect the leading order results of the variational calculation.

Equation~(\ref{eq:FGauss}) immediately leads to analytical forms for quantities of interest, including the confinement potentials,
\begin{eqnarray}
V_\gamma (z) &=& -\frac{e\sigma a_\gamma \beta_\gamma}{\epsilon \sqrt{8\pi}}
\left( e^{-2z^2/a_\gamma^2}-1 \right)
\nonumber \\ && \hspace{.25in}
-\frac{e\sigma \beta_\gamma}{2\epsilon} |z| \,
\text{erf} \left(\sqrt{2}|z|/a_\gamma \right) , \label{eq:VGauss}
\end{eqnarray}
where $\text{erf}(x)$ is the error function.  Equations~(\ref{eq:EG}) and (\ref{eq:ED}) then reduce to
\begin{eqnarray}
E_\Gamma &=& \frac{\hbar^2}{2m_la_\Gamma^2}
-\frac{e\sigma}{\epsilon\sqrt{8\pi}} (1-\beta_\Gamma) \sqrt{a_\Gamma^2+a_\Delta^2}
\label{eq:EGGauss} \\ && \hspace{0in} \nonumber 
+\frac{e\sigma}{\epsilon \sqrt{8\pi}} \left[ \left( 1+\frac{1-\sqrt{2}}{2}\beta_\Gamma
\right)a_\Gamma+(1-\beta_\Gamma ) a_\Delta \right] ,
\nonumber \\  \hspace{0in} 
E_\Delta &=& \frac{\hbar^2}{2m_ta_\Delta^2}
-\frac{e\sigma}{\epsilon\sqrt{8\pi}} \beta_\Gamma \sqrt{a_\Gamma^2+a_\Delta^2} 
\label{eq:EDGauss} \\ && \hspace{0in} \nonumber 
+\frac{e\sigma}{\epsilon \sqrt{8\pi}} \left[ \left( \frac{3-\sqrt{2}}{2}+\frac{\sqrt{2}-1}{2}\beta_\Gamma
\right)a_\Delta+\beta_\Gamma a_\Gamma \right] ,
\end{eqnarray}
in terms of the variational parameters.

To complete the variational analysis, we must minimize the average electron energy with respect to the variational parameters.  Here, we take the slightly different approach of minimizing the band energies $E_\Gamma$ and $E_\Delta$, while introducing a level filling constraint from Eq.~(\ref{eq:be}).  The constrained problem is then converted to an unconstrained problem by employing a Lagrange multiplier $\lambda$.  The minimization statement becomes
\begin{equation}
\bm{\nabla} (E_\Gamma + \lambda g)=0 , \label{eq:nabla}
\end{equation}
where the derivative is taken with respect to variables $a_\Gamma$, $a_\Delta$, $\beta_\Gamma$, and $\lambda$, and we have defined
\begin{equation}
g=E_\Delta - E_\Gamma+\frac{\pi \sigma \hbar^2}{4e\sqrt{m_tm_l}}
-\beta_\Gamma \left( 1+\sqrt{\frac{m_t}{4m_l}}\right) \frac{\pi \sigma \hbar^2}{2em_t} .
\end{equation}
Note that the $\lambda$ derivative is equivalent to setting $g=0$.  

Eliminating the Lagrange multiplier from Eq.~(\ref{eq:nabla}) leads to a system of three equations
\begin{gather}
g=0 \label{eq:Apreg} , \\
\frac{\partial E_\Gamma}{\partial a_\Gamma}
\, \frac{\partial E_\Delta}{\partial a_\Delta}
=\frac{\partial E_\Gamma}{\partial a_\Delta} \, \frac{\partial E_\Delta}{\partial a_\Gamma} ,
\label{eq:Apre} \\
\frac{\partial E_\Gamma}{\partial a_\Gamma} 
 \left[ \frac{\partial E_\Delta}{\partial \beta_\Gamma}
-\left( 1+\sqrt{\frac{m_t}{4m_l}} \right) \frac{\pi \sigma \hbar^2}{2em_t} \right]
=\frac{\partial E_\Gamma}{\partial \beta_\Gamma} \, \frac{\partial E_\Delta}{\partial a_\Gamma} ,
\label{eq:Apreb}
\end{gather}
which may be solved to obtain estimates for the variational parameters.  We do not report on such an analysis (yet), since Eqs.~(\ref{eq:Apreg})-(\ref{eq:Apreb}) cannot be solved exactly by analytical methods, and since we will perform a more rigorous numerical analysis later, which includes other contributions to the physics.  The variational formulation, however, leads immediately to an important scaling theory, which we discuss now.

\subsection{Scaling Theory} \label{sec:scaling}
Our simple variational theory describes the main portion of the wavefunction envelopes correctly, and should therefore capture the leading order physics of the $\delta$-doping problem.  Based on this statement, we may draw some very general conclusions, which can be expressed in terms of a scaling theory.  Of particular interest, the scaling theory captures the principal dependence of various quantities of interest, regarding the doping density $\sigma$.

The main expressions entering the variational procedure are given in Eqs.~(\ref{eq:EGGauss}), (\ref{eq:EDGauss}), and (\ref{eq:Apreg}) [or (\ref{eq:be})].  We can reformulate Eqs.~(\ref{eq:EGGauss}) and (\ref{eq:EDGauss}) in terms of dimensionless variables as follows:
\begin{gather}
a_\gamma = \left( \frac{\hbar^2\epsilon\sqrt{8\pi}}{em_0\sigma} \right)^{1/3} \tilde{a}_\gamma ,
\label{eq:dima} \\
E_\gamma = \left( \frac{e^2\sigma^2\hbar^2}{8\pi m_0\epsilon^2} \right)^{1/3} \tilde{E}_\gamma ,
\label{eq:dimE}
\end{gather}
where the quantities with tildes are dimensionless, and $m_0$ can be taken as the bare electron mass.  Recall here that $E_\gamma$ refers to the ground state confinement energy of the $\Gamma$ or $\Delta$ band, as measured from the bottom of the confinement potential.  When the energy is normalized in this way, $E_\gamma$ is strictly positive.

Since $(m_t/m_0)$, $(m_l/m_0)$ and $\beta_\Gamma$ are all of order unity, we expect that $\tilde{a}_\gamma$ and $\tilde{E}_\gamma$ should also be of order unity.  Indeed, we may go beyond the variational calculation described above to obtain more rigorous numerical estimates (described below), for the case of 1/4~ML filling.  The results are shown in Table~\ref{tab:bvalues}.  Note that these estimates could be improved by using results from rigorous microscopic calculations, or from experiments.  The scaling theory itself, however, would remain unaffected.

Based on Eqs.~(\ref{eq:dima}) and (\ref{eq:dimE}), we can deduce the scaling behaviors for other quantities of interest.  For example, from Eq.~(\ref{eq:be}) we obtain the relative filling fractions
\begin{gather}
\beta_\Gamma \simeq 0.19+\tilde{b} \left( \frac{e^5m_0^2}{\sigma\hbar^4\pi^4\epsilon^2} \right)^{1/3}  , \\
\beta_\Delta \simeq 0.81-\tilde{b} \left( \frac{e^5m_0^2}{\sigma\hbar^4\pi^4\epsilon^2} \right)^{1/3} .
\label{eq:betaD}
\end{gather}

In the Hartree theory, the electrostatic potential is defined as $V(z)=V_i(z)+V_\Gamma(z)+V_\Delta(z)$.  The depth of the confinement potential, $V_0=V(\infty)-V(0)$, plays an important role for the quantum theory.  Within the variational approach described above, this quantity can be expressed as
\begin{equation}
V_0=\frac{e\sigma}{\epsilon \sqrt{8\pi}} \left( a_\Gamma\beta_\Gamma+a_\Delta\beta_\Delta \right) . \label{eq:V0}
\end{equation}

\begin{table}[t]
  \begin{center}
    \begin{tabular}{ccc}
      \hline \hline
      Parameter	& 	value	 &  Eq. \# \\  \hline
	$\tilde{a}_\Gamma$ & 1.23 & \ref{eq:dima} \\
	$\tilde{a}_\Delta$ & 2.49 & \ref{eq:dima} \\
	$\tilde{E}_\Gamma$ & 1.34 & \ref{eq:dimE} \\
	$\tilde{E}_\Delta$ & 2.13 & \ref{eq:dimE} \\
	$\tilde{b}$ & 0.13 & \ref{eq:betaD} \\
	$\tilde{v}$ & 0.89 & \ref{eq:V0} \\
      \hline \hline
    \end{tabular}
  \end{center}
  \caption{Dimensionless parameters appearing in the scaling theory.  All scaling parameters were determined numerically, from the case of 1/4~ML doping, as described in Sec. \ref{sec:methods}.}
  \label{tab:bvalues}
\end{table}

To conclude this section, we note that the scaling theory breaks down outside a regime of validity.  In the present analysis, we have assumed a jellium model for the doping.  In the low density limit, however, the jellium model breaks down when the average dopant separation $\sqrt{e/\pi \sigma}$ approaches the characteristic effective mass length scale, $\text{min} [a_\Gamma,a_\Delta]$.  Within the scaling theory, we can estimate this breakdown density as 1/18~ML.  
In the high density limit, it is important to note that we have only included the $\Gamma$ and $\Delta$ bands in the present analysis.  For densities larger than 1/4~ML, the filling of additional bands becomes important.  This can easily be accomplished and incorporated into the present formalism although it lies outside the scope of the present work.


\subsection{Valley Splitting}
\label{sec:VS}

As discussed in Sec.~\ref{sec:2DEMT}, the combination of inhomogeneous (vertical) confinement and effective mass anisotropy lifts the degeneracy of the bulk bands.  The resulting splittings can be quite large.  Remaining degeneracies are lifted when the confinement potential is sharp.  For patterned devices, the valley splitting in the $\Delta$ band is extremely small, due to weak lateral confinement.\cite{martin}  In this section, we focus on the coupling between the $z$ valleys, due to the sharp $\delta$-doping potential, which splits the $\Gamma$ band to form $\Gamma_1$ and $\Gamma_2$ bands.

The envelope function equations (\ref{eq:epG}) and (\ref{eq:epD}) do not explicitly take into account the fact that the envelopes are formed from Bloch states within a given valley.  We can account for this translation in the Brillouin zone in a simple way, by introducing an overall phase factor.\cite{Kohn}  In the absence of valley coupling, we may therefore define the $z$-valley basis as follows:\cite{Fritzsche62,Twose62,friesen07}
\begin{equation}
f_\pm (z)=e^{\pm ik_0z}F_\Gamma(z) .
\end{equation}
Here, $F_\Gamma(z)$ is obtained from Eq.~(\ref{eq:epG}), and the resulting basis states $f_\pm (z)$ are degenerate.  Valley coupling can be treated perturbatively in the same basis, through the Hamiltonian
\begin{equation}
H_\Gamma=\begin{pmatrix} \varepsilon_\Gamma && V_\text{VO} 
\\ V_\text{VO}^* && \varepsilon_\Gamma \end{pmatrix} .
\end{equation}
There are two types of contributions which enter the valley-orbit coupling term $V_\text{VO}$.  The first type cannot be described within an EMT.  These include the so-called central cell corrections, which arise due to core electrons,\cite{YuBook} as well as discretization effects associated with the crystal lattice.\cite{Chutia2008}  The latter contributions are fairly weak.  Central cell corrections are also weak for shallow donors such as Si:P.  Indeed, for isolated donors, the main contributions to valley-orbit coupling may be treated effectively using methods similar to those described below.\cite{Friesen05}  For simplicity, we therefore ignore non-EMT corrections here.

Instead, we focus on valley-orbit couplings, which may be treated perturbatively within the EMT.  They are defined as
\begin{equation}
V_\text{VO} = \langle + | V | - \rangle = \int_{-\infty}^\infty V(z) F_\Gamma^2(z)
e^{-2ik_0z}dz , \label{eq:VVO}
\end{equation}
where $V(z)$ is the electrostatic confinement potential.  The valley-split single-electron energy levels are then given by $\varepsilon_1=\varepsilon_\Gamma-|V_\text{VO}|$ and $\varepsilon_2=\varepsilon_\Gamma+|V_\text{VO}|$, while the valley splitting is given by $2|V_\text{VO}|$.  Similarly, the individual band minima are given by 
\begin{equation}
E_1=E_\Gamma- |V_\text{VO}|, \quad\quad E_2=E_\Gamma+ |V_\text{VO}| . \label{eq:E12}
\end{equation}

The self-consistent numerical solutions, described below, are unaffected by valley splitting.  This becomes clear if we note that while the $\Gamma_1$ and $\Gamma_2$ bands fill differently, the electrostatic equations depend only on the combined filling factor, $\beta_\Gamma=\beta_1+\beta_2$.  Likewise, the quantum mechanical envelope function equations are identical for $\Gamma_1$ and $\Gamma_2$.  We may therefore solve for the energies $E_\Gamma$ and $E_\Delta$ and the fillings $\beta_\Gamma$ and $\beta_\Delta$, as we did previously, while computing the perturbations due to valley splitting \emph{a posteriori}.  After solving for the envelope functions, $V_\text{VO}$ is determined from Eq.~(\ref{eq:VVO}).  The band minima are then obtained from Eq.~(\ref{eq:E12}), while the fillings are obtained from
\begin{equation}
\beta_{1,2}=\frac{\beta_\Gamma}{2}\pm \frac{em_t|V_\text{VO}|}{\pi\sigma\hbar^2} .
\label{eq:betaav}
\end{equation}

It is interesting to analyze the scaling behavior of the valley splitting, since energy splittings can be measured experimentally, via spectroscopy techniques.
 ~Since $k_0\sim 1/a$, Eq.~(\ref{eq:VVO}) may be regarded as an integral transform that picks out the Fourier components in $V(z)F_\Gamma^2(z)$ with very short wavelengths.  The predominant feature at short wavelengths is the sharply peaked confinement potential at $z=0$.  The more slowly varying features occurring away from $z=0$ effectively cancel out, and do not contribute greatly to the integral.  (In some cases, a sharp variation of the wavefunction can also contribute to $V_\text{VO}$,\cite{Saraiva09} however, we ignore this possibility in the simple estimate presented here.)  We may therefore approximate $V_\text{VO}$ by truncating the integration range in Eq.~(\ref{eq:VVO}) to a single oscillation of the exponential, from $z=-\pi/2k_0$ to $\pi/2k_0$.  Over this range, we can approximate $V(z)F_\Gamma^2(z)\simeq e\sigma|z|/\sqrt{2\pi}\epsilon a_\Gamma$, leading to the following estimate for the valley splitting:
\begin{equation}
2|V_\text{VO}| \sim \frac{e\sigma}{\sqrt{2\pi}\, \epsilon a_\Gamma k_0^2} .
\end{equation}
Despite the obvious roughness of this estimate, we expect it to encompass the leading order contributions to the scaling theory, which we find to be
\begin{equation}
2|V_\text{VO}| = \left( \frac{m_0\, e^4\sigma^4}{8\pi^2\hbar^2\epsilon^4k_0^6} \right)^{1/3} 
\tilde{v} . \label{eq:vtilde}
\end{equation}
Thus, we note that the $\Gamma_1$-$\Gamma_2$ splitting exhibits a much stronger dependence on the doping density $\sigma$ than does the $\Gamma_1$-$\Delta$ splitting, as borne out by the numerical analysis, described below.

\begin{figure}[t] 
  \centering
  \includegraphics[width=3.1in,keepaspectratio]{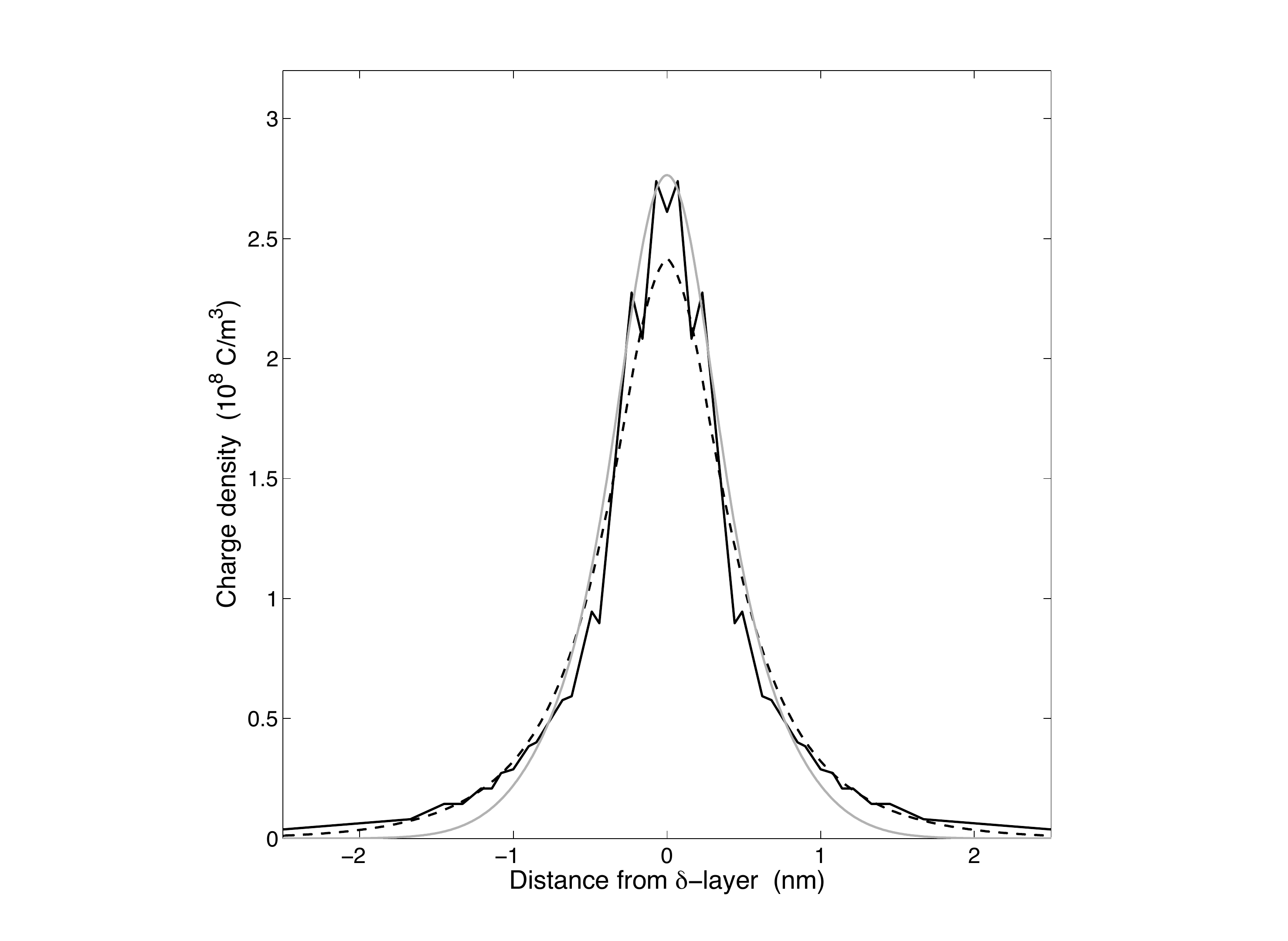}
  \caption{Electronic charge density for the case of 1/4~ML doping in a Si:P $\delta$-layer.  Black solid line:  density functional theory from Ref.~\onlinecite{oliver}.  The density functional results exhibit small oscillations, which arise due to perfect ordering in the 2D dopant array.  Gray line: variational calculation, from Sec.~\ref{sec:variational}.  Dashed black line:  self-consistent numerical solutions for the wavefunction envelope, as described in Appendix~\ref{sec:1Dsim}.  Note that we have assumed a smooth lateral distribution for the dopants, via the jellium model.  The resulting effective mass solutions do not oscillate.  Also note that the Gaussian variational solution accurately represents the full numerical results, except in the tail region.}
  \label{fig:wavefn}
\end{figure}


\subsection{Exchange and Correlation}
\label{sec:LDA}
Self-consistent many-body effects were included in the variational calculations of Sec.~\ref{sec:variational}.  There, we employed a Hartree theory, in order to simplify our analytical calculations.  For more accurate numerical results, it is important to also include exchange (X) and correlation (C) effects.  We do this here, using the local density approximation (LDA).\cite{Kohn65a}  Although we are primarily interested in higher densities, it is well known that exchange and correlation effects are most significant at lower densities.  For completeness, we will therefore study a wide range of densities, over which the LDA must remain valid.  Typically, this requires specially developed parameterizations.\cite{Vosko80,Perdew81,Perdew92}  Here, we follow the method of Scolfaro \textit{et al.}\cite{Scolfaro94} and Rodriguez-Vargas \& Gaggero-Sager,\cite{Rodriguez-Vargas06} where the LDA is applied to the EMT equations, rather than at the atomistic level, although we use a more recent and accurate density functional parameterization developed by Perdew and Wang \cite{Perdew92}.  The final outcome is a pair of new potential terms, $V_\text{X}(z)$ and $V_\text{C}(z)$, which we add to the total electron confinement potential $V(z)$:
\begin{equation}
V^{'}=V+V_{\text{X}}+V_{\text{C}}.
\end{equation}

Following Perdew and Wang, the exchange and correlation potentials may be approximated as
\begin{eqnarray}
V_\text{X} (z) &=& \frac{\partial (n \varepsilon_x)}{\partial n }
=-\left[ \frac{m^*e^4}{(4\pi\epsilon \hbar)^2}\right] 
\left( \frac{9\pi}{4} \right)^{1/3} \frac{1}{\pi r_s} ,  \label{eq:Vx}      \\ 
V_\text{C}(z) &=& \frac{\partial (n \varepsilon_c)}{\partial n }
\\ \nonumber 
&=& -2A\left[ \frac{m^*e^4}{(4\pi\epsilon \hbar)^2}\right] \biggl\{ 
\left(1+\frac{2\alpha_1r_s}{3} \right) 
\ln \left[ 1+\frac{1}{2Af} \right]
\\ \nonumber && \hspace{.5in}
+\frac{(1+\alpha_1r_s)}{3} \, 
\frac{f'}{f(1+2Af)} \biggr\} ,
\end{eqnarray}
where the 3D particle density is given by $n(z)=\rho(z)/e$, and we define
\begin{gather}
r_s (n) = \left( \frac{4\pi a^{*3}n}{3} \right)^{-1/3}, \\
f(r_{s})=b_{1}r_{s}^{1/2}+b_{2}r_{s}+b_{3}r_{s}^{3/2}+b_{4}r_{s}^{2} .
\end{gather}
Following Refs.~\onlinecite{Scolfaro94} and \onlinecite{Rodriguez-Vargas06}, we adopt a geometrically averaged effective mass, $m^*=(m_t^2m_l)^{1/3}$, and an averaged effective Bohr radius, $a^*=(4\pi\epsilon \hbar^2/m^*e^2)$.  The parameterization constants we use are appropriate in the absence of spin polarization: $A=0.031091$, $\alpha_1=0.21370$, $b_1=7.5957$, $b_2=3.5876$, $b_3=1.6382$, and $b_4=0.49294$.  

We can estimate the magnitude of the exchange and correlation terms.  We specifically consider the case of 1/4~ML $\delta$-doping.  Because of the high doping density, we find that the dimensionless electron separation length  $r_s$ ranges from about 0.6 at the center of the $\delta$-doping layer to $\infty$ far away from the doping plane.  In the high density region, which is our main interest here, the exchange potential dominates over the correlation potential.  We can estimate its depth directly from Eq.~(\ref{eq:Vx}), finding that $V_{\text{X}0} \simeq 70$~meV.  This may be compared to the electrostatic potential depth in Eq.~(\ref{eq:V0}), which we find from numerical calculations to be $V_{0} \simeq 670$~meV.

In the numerical calculations discussed below, we solve the $\delta$-doping problem using the 3D~EMT, including valley splitting, exchange and correlation effects, as outlined in Appendix~\ref{sec:1Dsim}.  Overall, we find that exchange and correlation have relatively small effects on the population of the bands or on the valley-splitting.  The $z$-confinement of the electrons, however, is enhanced, particularly in the $\Delta$-band.  This is to be expected, since the exchange and correlation interactions both deepen the potential well, particularly near the doping layer, where the electron density is highest.  Since the $\Delta$ wavefunctions still spread out well beyond the doping plane, the overall effect of exchange and correlation on the $\Gamma$-$\Delta$ splitting is fairly weak.  The effect on the $\Gamma_{1}$-$\Gamma_{2}$ valley splitting is stronger, however, as it is directly related to the sharpness of the confinement potential.  Additional discussion of the exchange and correlation contributions to $\delta$-doping is presented in Appendix~\ref{sec:XC}.

\section{Results and Comparison}
\label{sec:methods}

In this section, we describe the numerical results of our 3D-EMT description of $\delta$-doping in Si:P, with details given in the Appendices.  We also compare our results to other reports in the literature, mainly based on density functional theory.  The closest points of comparisons include the planar Wannier orbital (PWO) method of Qian \emph{et al.},\cite{qian05} and the full density functional calculations of Carter \emph{et al.},\cite{Carter2011} using the ``single-zeta plus polarization" (SZP) basis set.  Both of these techniques may be modified to include disorder effects.  For the PWO method, this was accomplished using laterally averaged confinement potentials.  For the SZP method, quasi-random dopant arrays were considered, as well as ``mixed'' pseudopotentials, averaged over the atoms in the doping layer.  In the interests of brevity and generality, we compare explicitly to Ref. \onlinecite{qian05}, and also to the full \textit{ab initio} results of Ref. \onlinecite{oliver}.  Other abbreviations used hence are as follows:  our effective mass theory (EMT), the PWO method including short-ranged interactions between the dopants (PWOf),\cite{qian05} and the fully ordered (SZPo) vs. partially disordered dopant arrays (SZPd) discussed in Ref. \onlinecite{oliver}.

Before discussing our main numerical results, it is important to emphasize that the parabolic band structure assumed in the EMT (\emph{e.g.}, Fig.~\ref{fig:Energies}) is far more consistent with the highly disordered implementations of the density functional theory.  The small unit cells associated with dopant ordering generate effective terms in the Hamiltonian which couple the donor bands and lead to band structures that differ greatly from bulk silicon.\cite{oliver}  We therefore conjecture that the EMT based on the jellium donor model should be understood as a highly disordered model.  Dopant ordering, or any specific type of disorder, can be introduced into the EMT through additional modifications of the jellium model.  The smooth charge profiles obtained by EMT in Fig.~\ref{fig:wavefn} are also consistent with spatial averaging in the presence of disorder, as compared with the more oscillatory profile obtained from density functional theory for an ordered dopant array.

\begin{figure}[t] 
  \centering
  \includegraphics[width=3.1in,keepaspectratio]{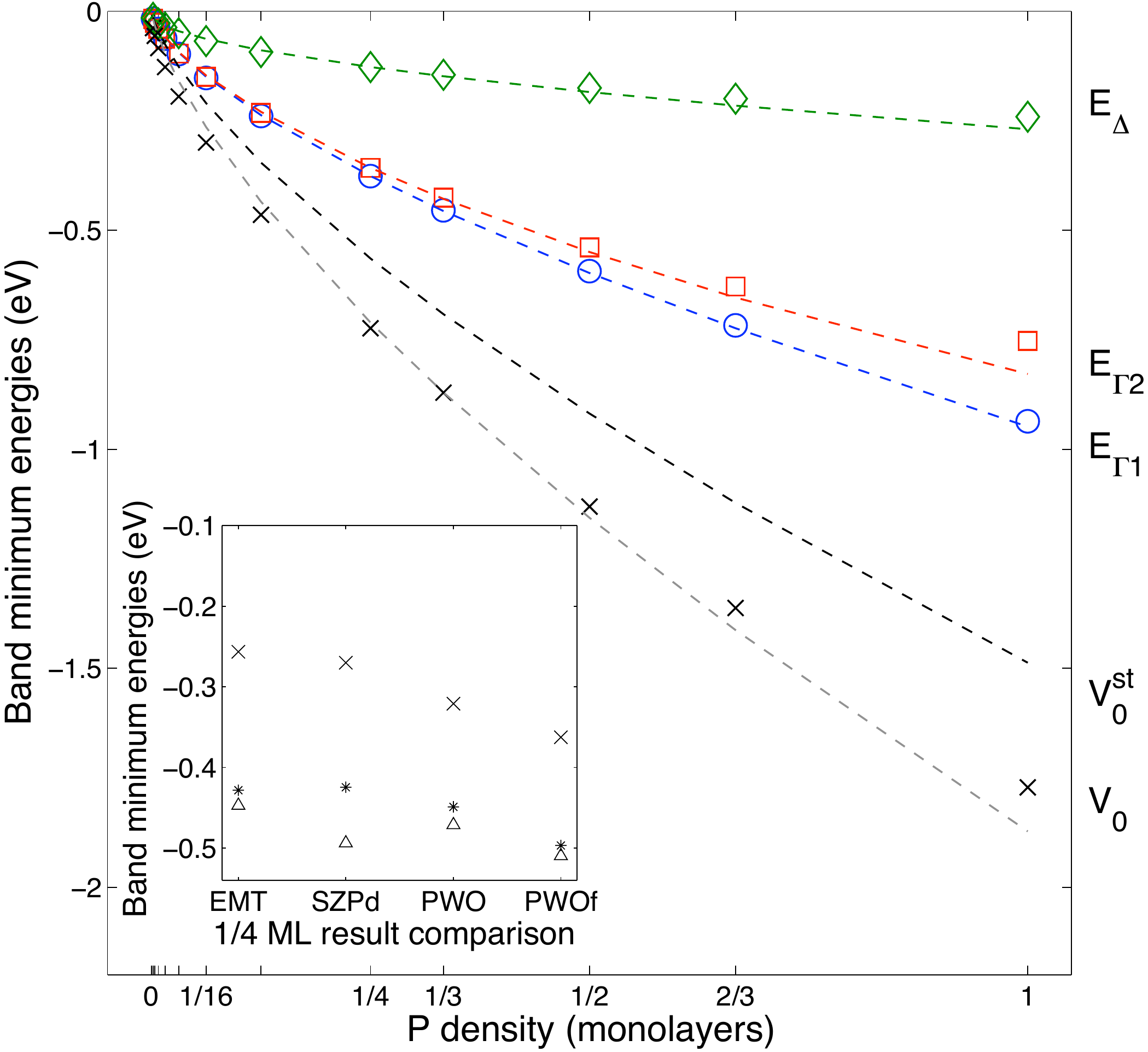}
  \caption{(Color online) 
  Minimum band energies relative to the bulk band minima for the $\Gamma_1$ band (blue circles), $\Gamma_2$ band (red squares) and $\Delta$ band (green diamonds).  Scaling theory fits are also shown as dashed lines.  The minima of the confinement potential, $V_{0}$, are shown as black crosses.  The corresponding scaling theory for the confinement minima V$_{0}^{\text{st}}$, based on derived parameters, is shown as a dashed black line (see main text).  A rescaled fitting is shown as a dashed grey line.  Inset shows a comparison of 1/4~ML results with Refs. \onlinecite{oliver} (SZPd) and \onlinecite{qian05} (PWO, PWOf) for the minima of the $\Gamma_{1}$ (triangles), $\Gamma_{2}$ (asterisks), and $\Delta$ (crosses) bands.  (See main text for abbreviations.)}
  \label{fig:Energies2}
\end{figure}

\begin{figure}[t] 
  \centering
  \includegraphics[width=3.1in,keepaspectratio]{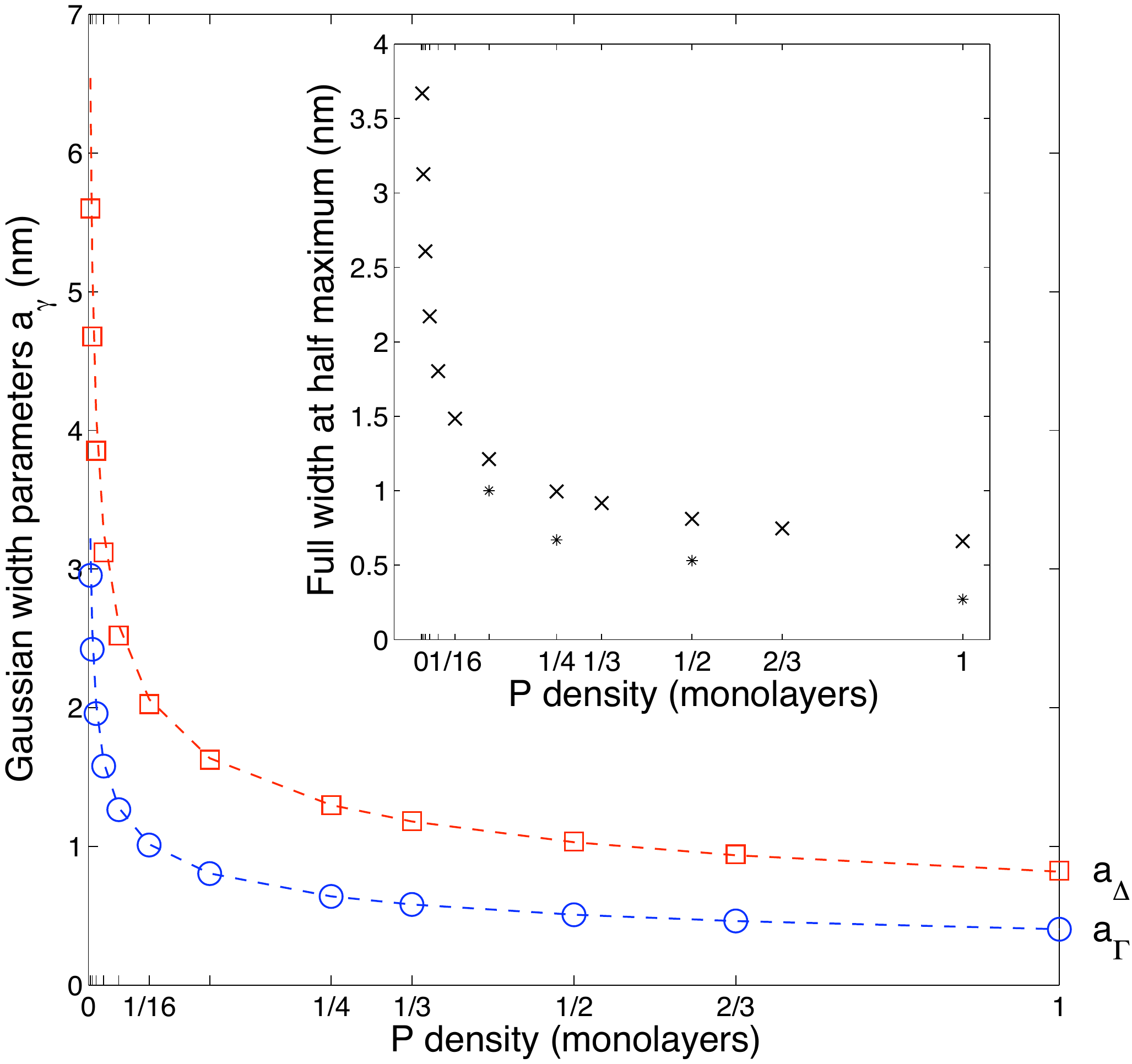}
  \caption{(Color online)
  Gaussian widths of the wavefunction envelopes: $a_{\Gamma}$ (blue circles), $a_{\Delta}$ (red squares).
  Scaling theory fits are shown as dashed curves.  Inset shows comparison of results for the full-width at half-maximum of the full electronic density: this work (crosses), SZPo results from Ref. \onlinecite{oliver} (asterisks).}
  \label{fig:Widths}
\end{figure}

\begin{figure}[t] 
  \centering
  \includegraphics[width=3.1in,keepaspectratio]{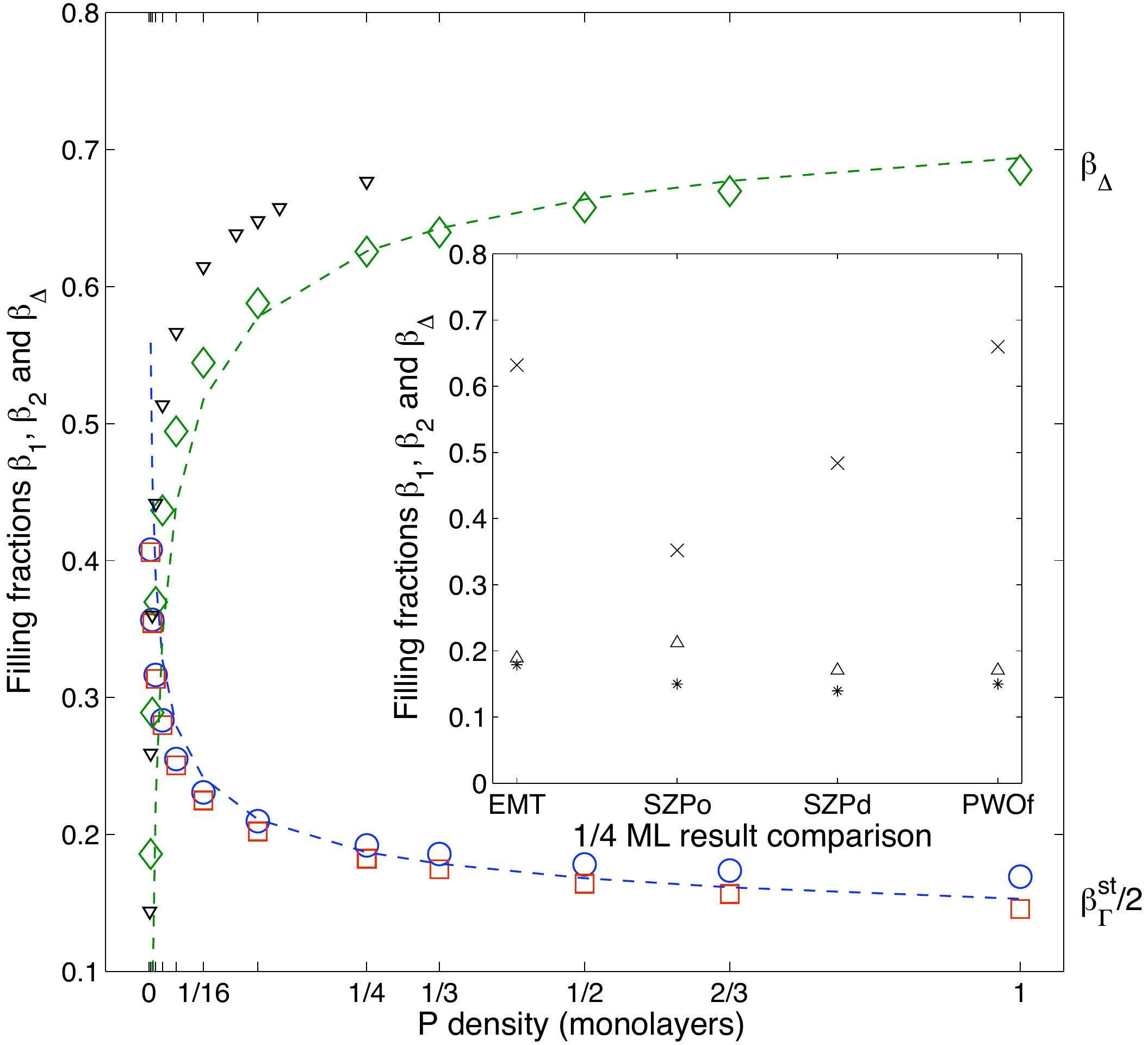}
  \caption{(Color online)
  Filling fractions for the different bands: $\beta_{1}$ (blue circles), $\beta_{2}$ (red squares), and $\beta_{\Delta}$ (green diamonds).  Scaling theory fits are shown as dashed curves (see main text); as discussed in Sec. \ref{sec:VS}, and in light of Eq. (\ref{eq:betaav}), half the scaling theory value for the $\Gamma$ filling fraction ($\beta^{\text{st}}_{\Gamma}/2$) is plotted. Black, downward-pointing triangles correspond to $\beta_{\Delta}$ results from Ref.~\onlinecite{qian05}.  Inset shows a comparison of $\beta_1$ (asterisks), $\beta_2$ (triangles), and $\beta_\Delta$ (crosses) for the case of 1/4~ML doping.  (See main text for abbreviations.)}
  \label{fig:betas}
\end{figure}

For the case of 1/4-ML doping, we can summarize our main EMT results as follows.  A fit of the numerical wavefunction envelopes to the Gaussian form used in the variational procedure of Sec.~\ref{sec:variational} gives the envelope widths $a_{\Gamma}=0.64$~nm and $a_{\Delta}=1.30$~nm.  The band filling parameters are given by $\beta_{1}=0.19$, $\beta_{2}=0.18$, and $\beta_{\Delta}=0.63$, while the valley splitting is given by $2|V_\text{VO}|=19$~meV.  
In Figs.~\ref{fig:Energies2}-\ref{fig:betas}, we plot our numerical results for the band minima $E_\Gamma$ and $E_\Delta$, the wavefunction widths (in terms of the Gaussian fitting parameters $a_\Gamma$ and $a_\Delta$), and the band filling fractions $\beta_\Gamma$ and $\beta_\Delta$, as a function of doping density.  The figure insets provide additional comparisons with the literature.  (In some cases, the values have been determined graphically, from published plots.)

We now discuss the main plots in Figs.~\ref{fig:Energies2}-\ref{fig:betas} in more detail.  In each plot, the markers represent numerical results obtained as a function of P doping density in the $\delta$ layer, while the curves reflect the corresponding scaling theory parameters given in Table ~\ref{tab:bvalues}.  Direct comparisons are made to Ref. \onlinecite{qian05} (a fully-disordered technique), and to Ref. \onlinecite{oliver} (a full \textit{ab initio} technique).

Figure \ref{fig:Energies2} shows the energies of the various band minima measured from the bottom of the bulk (undoped) Si conduction band.  In these calculations, we have not considered background dopants, so the bulk band minimum ($V=0$) corresponds to the asymptotic value of the confinement potential $V(z)$ far from the doping layer.  As the doping increases, $V(z)$ deepens significantly, dragging the confinement energy levels with it.  The valley splitting between the $\Gamma_{1}$ and $\Gamma_{2}$ band minima also increases quickly at higher densities.  The inset shows agreement between EMT and the SZPd method for the $\Delta$ and $\Gamma_2$ bands, and agreement with the PWO and PWOf methods for the valley splitting, for the 1/4 monolayer doping density case.

We observe that the scaling theory describes the location of the band minima quite well.  (The scaling of the valley splitting will be discussed below.)  It appears that the scaling theory is less accurate for the potential well minimum $V_0=V(\infty)-V(0)$.  In this case, the scaling form (dashed black line) was derived from Eqs.~(\ref{eq:dima})-(\ref{eq:V0}) using the parameters in Table~\ref{tab:bvalues}.  This discrepancy in the scaling theory is primarily due to the inclusion of correlation and exchange effects in the numerical solution, while they are absent from the discussion leading to the scaling theory.  Exchange and correlation both tend to deepen the confinement potential, and they both have their greatest (absolute) effect at the origin.  It is a sign of robustness of the scaling theory that such corrections can be accommodated by a simple adjustment of the scaling parameters, as indicated by the dotted black line.

The $a_\gamma$ parameters, or Gaussian widths for the wavefunctions, are shown in Fig.~\ref{fig:Widths}.  As expected, higher doping tends to enhance the electron confinement, despite the Coulomb repulsion between the larger number of electrons.  At low densities, the electron wavefunctions appear unphysically large, although our predictions for densities lower than 1/18~ML should not be compared directly to physical systems, as explained in Sec.~\ref{sec:scaling}, due to the breakdown of the jellium approximation.

The scaling theory provides an excellent description of the numerical results over the entire range of densities in Fig.~\ref{fig:Widths}.  Small inaccuracies of the scaling theory may be attributed to exchange and correlation effects, as discussed in Appendix~\ref{sec:XC}.  

We expect the filling fractions plotted in Fig.~\ref{fig:betas} to match the predictions of scaling theory quite well, because Eq.~(\ref{eq:be}) is exact for a parabolic band structure, and because the band minima are also well described by the scaling theory.  Note that we have plotted the results for $\beta_1$ and $\beta_2$ separately, and compared them to the scaling theory result for $\beta_\Gamma/2$, where $\beta_\Gamma=\beta_1+\beta_2$.  As before, the main deviations from the scaling theory occur at low densities where exchange and correlation effects are most important.

Because Eq.~(\ref{eq:be}) is generic, the scaling theory for $\beta_\gamma$ could also be applied to results from other methods, such as those in Ref.~\onlinecite{qian05}.  The latter ($\beta_{\Delta}$ values) are graphically estimated and shown in the main panel of Fig.~\ref{fig:betas}.  The asymptotic, high density values of $\beta_\gamma$ depend only on the effective mass, and  should be nearly identical to those calculated here.  The main deviations between the scaling theory and EMT at high densities arise because of small errors in the scaling theory for the quantity $(E_\Delta-E_\Gamma)$.  At low densities, the discrepancies are due to exchange and correlation effects.

For the 1/4 ML results shown in the inset of Fig. \ref{fig:betas} (graphically estimated from bandstructures in the relevant papers), our EMT results are most similar to PWOf.  For the SZP results, the disordered model (SZPd) is most similar to EMT.  This is consistent with our conjecture that the EMT provides a good description of the high disorder limit.  We also note that, using the definitions of $\beta$ in Eqs. (\ref{eq:D1})-(\ref{eq:DD}), it appears that the charge neutrality condition, Eq.~(\ref{eq:betasum2}), is not satisfied in either case (particularly in Ref. \onlinecite{oliver}) - though this is due, at least in part, to their inclusion of other bands, as discussed below.


\begin{figure}[t]
	\centering
	\includegraphics[width=2.5in,keepaspectratio]{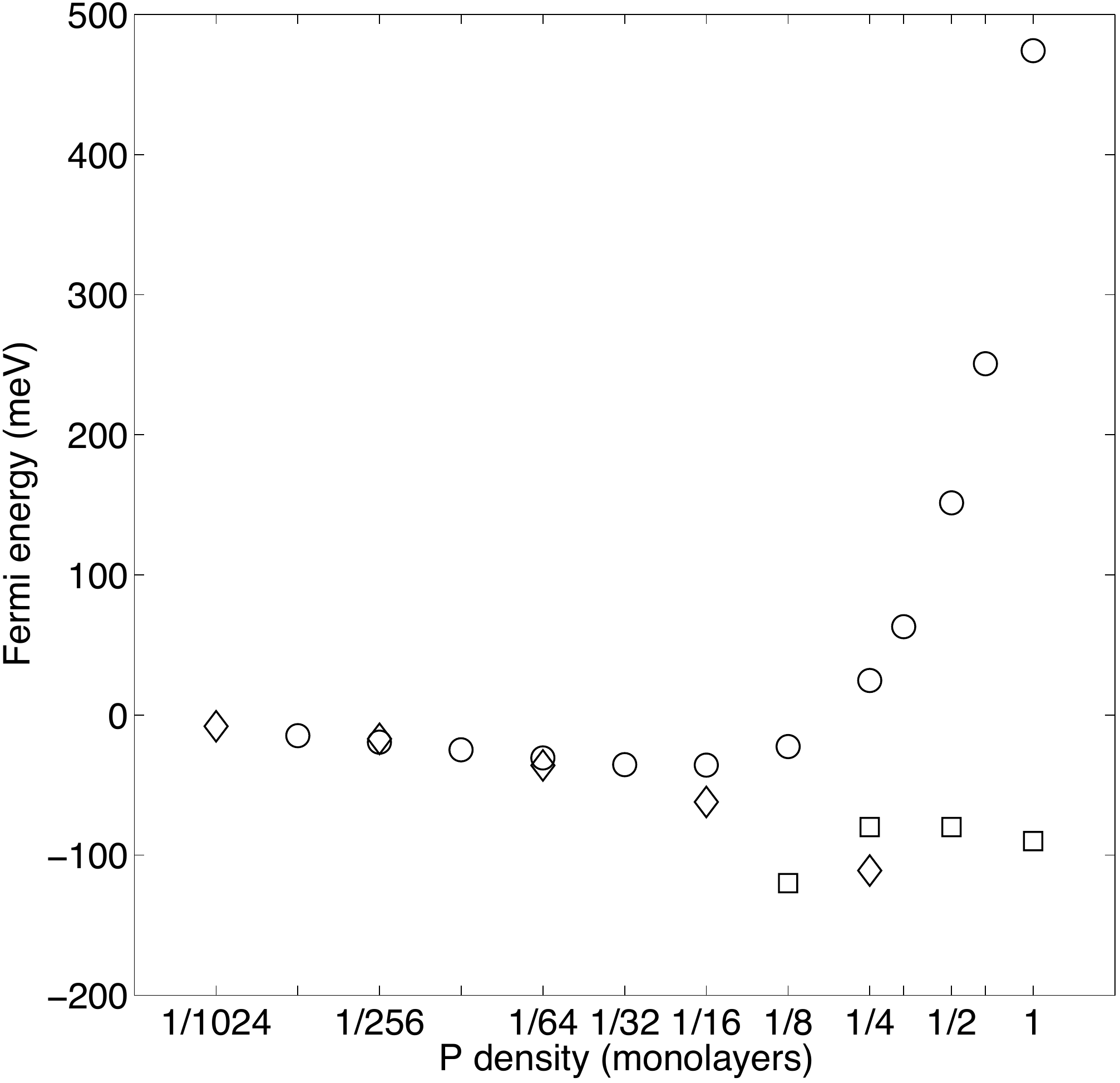}
  \caption{Comparison of Fermi levels obtained in this work (circles), PWO results from Ref.~\onlinecite{qian05} (diamonds), and SZPo results from Ref.~\onlinecite{oliver} (squares).}
  \label{fig:comp}
\end{figure}

Figure \ref{fig:comp} shows our calculated results for the Fermi energy, relative to the bulk conduction band minimum.  Some Fermi energies obtained by density functional methods are also shown.  Fermi levels are notoriously difficult to calculate accurately.  This is especially true for highly doped Si:P, due to the filling of multiple bands, and the fact that separate band minima must all be computed self-consistently.  The EMT results in Fig.~\ref{fig:comp} change sign, unphysically, near the 1/4~ML doping level.  This can be attributed to the conspicuous absence of the $1X$/$2X$ bands at this density, which we have chosen not to include in our model in light of the fact that the relevant filling fraction has been found to be less than 0.01\cite{qian05}.  Such high-lying bands would absorb high energy electrons, and would therefore lower the Fermi level as they begin to fill.  Although no experimental measurements of the Fermi level are available at the present time, preliminary tunneling experiments between nano-fabricated wires suggest a Fermi energy of about -20~meV for the case of 1/4~ML doping.\cite{Wilson}  We also direct the reader to further discussion in Appendix~\ref{sec:XC}.


\begin{figure}[t] 
  \centering
  \includegraphics[width=2.5in,keepaspectratio]{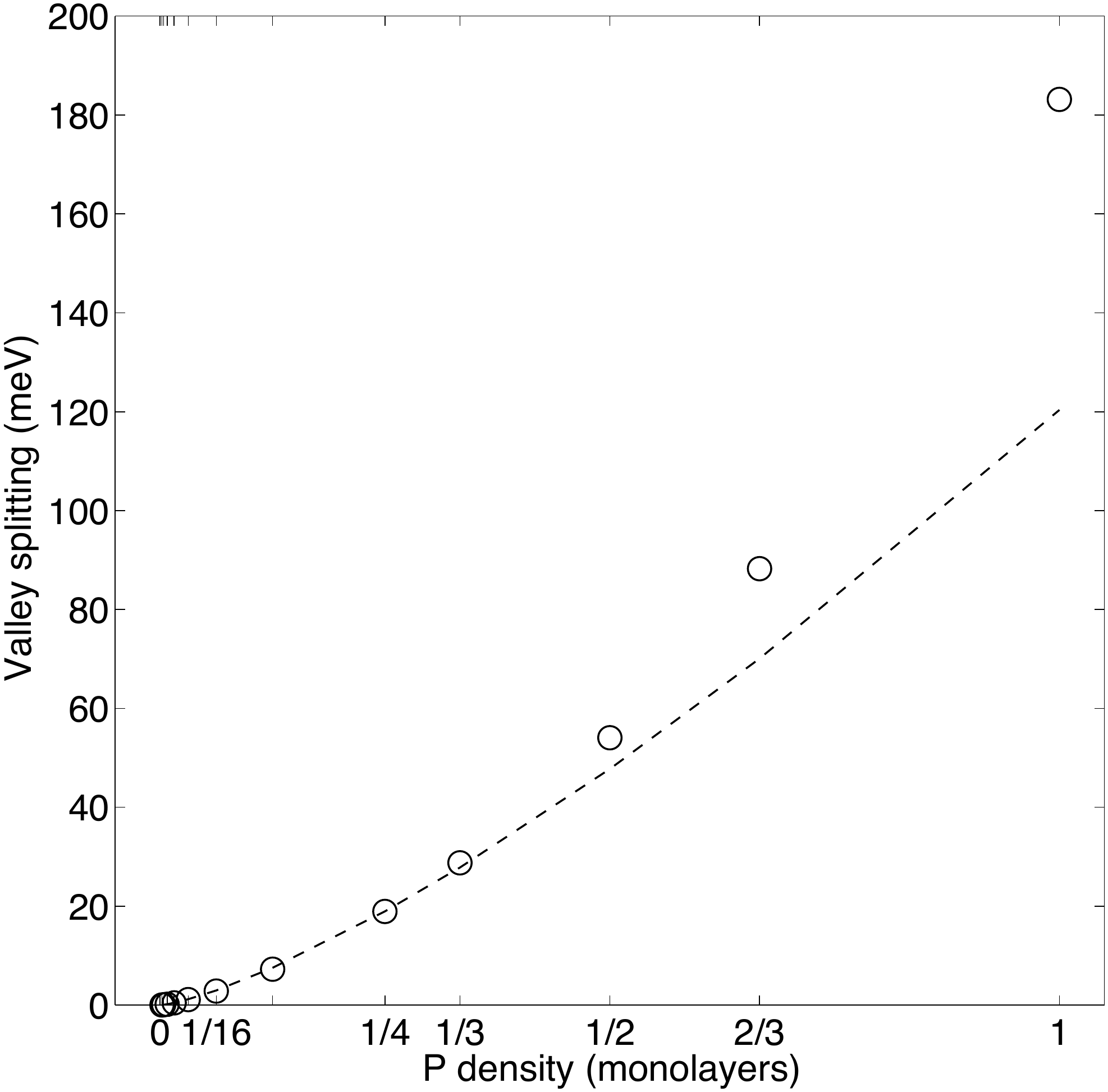}
  \caption{Numerical solutions for valley splitting between the $\Gamma_{1}$ and $\Gamma_{2}$ bands (circles), with scaling theory fit (dashed line).}
  \label{fig:VS}
\end{figure}

Finally, we plot our results for the valley splitting between the $\Gamma_1$ and $\Gamma_2$ band minima in Fig.~\ref{fig:VS}.  The valley splitting varies by several orders of magnitude over this density range.  Carter \textit{et al.} have noted that the valley splitting is particularly sensitive to the disorder model used in the calculations, with ordered dopants typically leading to larger valley splittings.\cite{Carter2011}

The scaling results shown in Fig.~\ref{fig:VS} provide an excellent representation of our numerical solutions.  It is interesting to note that, once again, the deviations are due to exchange and correlation effects (and partially to the truncation of the integral in Eq. \ref{eq:VVO} and subsequent linear approximation to the potential in this region).  In this case, however, the deviations are most evident at \emph{high densities}.  This occurs because exchange and correlation deepen the confinement potential at high densities, while the sharpness of the confinement potential provides the main contribution to the valley splitting.

%

\section{Summary and conclusions}
\label{sec:conc}
In this paper, we developed an effective mass theory for high density $\delta$-doped Si:P, and we argued that the model is consistent with the limit of high disorder.  The method was applied to study infinite planes of Si:P.  First a variational model was solved, which provided simple analytical results and demonstrated a remarkable agreement with density functional theories for very few assumptions.  Second, a more comprehensive numerical model was solved, including exchange and correlation effects, and valley splitting between the $\Gamma_1$ and $\Gamma_2$ bands.  Self-consistent solutions were obtained for systems comprised of $\delta$-layers with P densities ranging from $1/512$ to 1 monolayer.  

In our model the inclusion of valley-splitting in the self-consistent description has no effect on the extent of the donor wavefunctions, or on the relative populations of the $\Gamma$ and $\Delta$ bands.  Splitting between the $\Gamma_{1}$ and $\Gamma_{2}$ band minima naturally produces a difference in the filling of those bands.  Similar to Ref.~\onlinecite{qian05}, we find that electrons are mainly concentrated in the $\Gamma$-bands at low densities, while nearly two thirds of the electrons shift to the $\Delta$ band for densities approaching 1/4~ML.  We predict that the $\Delta$ filling should approach $70$-$80\%$ at higher densities, although such densities are not easily achieved in physical systems.

Since effective mass calculations mainly involve solving for coarse-grained envelope functions, they can be implemented much more efficiently than \textit{ab initio} methods, such as density functional theory, or tight-binding.  Speed comes at a price, however, since the technique relies upon accurate input data, including the anisotropic effective masses for the conduction electrons, the dielectric constant of the host material, and knowledge of the underlying bulk band structure.  The effective mass method may also have limited applicability at low and high densities.  In the first case, the jellium approximation breaks down below the metal-insulator transition.  In the second case, the $\Gamma$ and $\Delta$ bands tend to over-fill above 1/4~ML doping density, due to the absence of additional bands in the present theory.  Fortunately, the applicable range includes most problems of current experimental interest.  One of the main attractions of the effective mass theory is its versatility and its potential for treating complex problems of current interest for devices.\cite{martin}  The method is easily extended to higher dimension and large-scale geometries, which are typically out of reach for \emph{ab initio} techniques.

It is perhaps surprising that a minimal model like the effective mass theory could provide such a reasonable account of the broad range of physics in these $\delta$-layers.  The power of such a simple representation has been well-illustrated in the preceding sections, suggesting that the inputs to the effective mass theory capture the main physics in this problem.  Our results are well matched to the predictions not only of Qian \textit{et al.},\cite{qian05}\ whose technique incorporates disorder in a similar manner as ours, but also to Carter \textit{et al.},\cite{oliver} who utilize quasi-disordered dopant arrays.  Our calculated valley-splitting agrees with Ref. \onlinecite{qian05}, and our $\Gamma_{2}$-$\Delta$ band splitting and binding energies agree well with Ref.~\onlinecite{oliver}.  Our results for the band fillings compare well to other values in the literature; in particular to the fully-disordered results of Ref. \onlinecite{qian05}.

Finally, we have condensed our effective mass results into a scaling theory, which may represent the simplest and most far-reaching outcome of the theory of infinite Si:P planes.  The scaling theory reproduces our numerical results very well up to high doping densities of order 1/2~ML, and it enables analytical calculations of various physical quantities, as a function of the doping density.  Examples include the band energies, $E_{\gamma}\sim\sigma^{2/3}$, the Gaussian widths of the wavefunctions, $a_{\gamma}\sim\sigma^{-1/3}$, the depth of the confinement potential, $V_0\sim\sigma$, and the valley splitting, $2V_\text{VO}\sim \sigma^{4/3}$.

\section*{Acknowledgments}
M.\ F.\ gratefully acknowledges support from ARO and LPS (Grant No.\ W911NF-08-1-0482), NSF (Grant No.\ DMR-0805045), the University of New South Wales, where some of this work was completed, and is indebted to O.\ Warschkow for many helpful discussions.  D.\ W.\ D., L.\ C.\ L.\ H., and M.\ Y.\ S.\ are members of the Australian Research Council Centre of Excellence for Quantum Computation and Communication Technology (project number CE110001027), and acknowledge support from the US National Security Agency and the US Army Research Office under contract number W911NF-08-1-0527.  D.\ W.\ D. gratefully acknowledges support from the University of Melbourne (Overseas Research Experience Scholarship), and the University of Wisconsin, Madison, where some of this work was completed.

\appendix

\section{1D Numerical Procedure}
\label{sec:1Dsim}

In this appendix, we obtain full numerical eigenvalue solutions of Eqs.~(\ref{eq:epG}) and (\ref{eq:epD}), in the presence of correlation and exchange.  In contrast with the variational calculations of Sec. \ref{sec:variational}, we do not fix the value of $\beta_{\Gamma}$.  The method we use is iterative.  At each stage of the procedure, a new density profile is obtained.  The density is used to compute the electrostatic potential, which is used to compute a new density profile, and so on.  Self-consistency is accomplished by introducing a new constraint:
\begin{equation}
h=\int [n_{\alpha+1}(z)-n_\alpha(z)]^2 dz =0 . \label{eq:hconstraint}
\end{equation}
Here, $\alpha$ and $\alpha+1$ indicate the iteration number.  In other words, the density should not change from iteration to iteration once convergence is achieved.  

The self-consistency constraint of Eq.~(\ref{eq:hconstraint}) is well-defined and can be applied to density approximations involving many parameters.  For example, the density could be defined spatially as $n_i=n(z_i)$, where $i$ is now a spatial index (not the iteration index, $\alpha$).  In this case, the self-consistency procedure is high-dimensional, involving many independent parameters.  To simplify our analysis, we will calculate the electrostatic potential using Gaussian density profiles, similar to those in Sec. \ref{sec:variational}.  From Fig.~\ref{fig:wavefn}, we see that such Gaussian forms provide a very reasonable estimate for the density, and can be immediately integrated to obtain Hartree potentials, as in Eq.~(\ref{eq:VGauss}).  Indeed, by fitting Gaussian forms to ``exact" results for $F_\Gamma$ and $F_\Delta$, obtained by finite element methods, we obtain density approximations that are much more accurate than the variational approximation shown in Fig.~\ref{fig:wavefn}.  

Our self-consistent method is then expressed as follows:  (i) provide a Gaussian estimate for the density profile in iteration $\alpha$, (ii) incorporate the corresponding Hartree, exchange and correlation potentials into a finite element Schr\"{o}dinger solver, (iii) fit the resulting eigenfunctions $F_\Gamma$ and $F_\Delta$ to Gaussian forms, to be used in iteration $\alpha+1$, (iv) repeat until convergence is achieved, by minimizing the constraint of Eq.~(\ref{eq:hconstraint}), where $n_\alpha$ refers to the Gaussian forms.  The minimization step (\textit{i.e.}, the constraint) is multidimensional in the Gaussian parameters $a_\Gamma$ and $a_\Delta$, and can be accomplished using the BFGS method.\cite{Broyden70,Fletcher70,Goldfarb70,Shanno70}  Note that for a more accurate result, the wavefunction could be expanded in a larger basis of Gaussian functions.  Minimization of such a parameterization would still be accomplished more efficiently than for the spatial parameterization, $\{ n_i \}$.

It is possible to incorporate the self-consistency constraint of Eq.~(\ref{eq:hconstraint}) into the variational construction of Eq.~(\ref{eq:nabla}).  The quantity to be
minimized would then be $f=E_\Gamma+\lambda g + \mu h$.  In such an approach, $h=0$ would never be satisfied until convergence is achieved.  We have found, however, that the extended parameter space associated with allowing $h\neq 0$ introduces new local minima, which are difficult to avoid.  We therefore employ a different method, as described below.

The quantum mechanical problem involves two Schr\"{o}dinger equations, (\ref{eq:epG}) and (\ref{eq:epD}), whose coupling, through the electrostatic potential $V(z)$, is fully specified by the parameter $\beta_{\Gamma}$.  Consequently, the parameters $a_\Gamma$ and $a_\Delta$ are completely determined by $\beta_{\Gamma}$, and do not depend on parameters used in the variational approach, such as $\lambda$.  This statement remains true for more complex geometries, such as the 2D geometries considered in Ref. \onlinecite{martin}. 
In such cases, many parameters may be required to fully describe the wavefunctions, although self-consistency still depends only on $\beta_{\Gamma}$.  We can therefore use a numerical approach where the Schr\"{o}dinger equations are solved self-consistently for a fixed value of $\beta_{\Gamma}$.  This one parameter is then varied, in order to satisfy the Fermi level constraint, $g=0$.  The latter problem is simple, and can be accomplished using a Newton-Raphson method.

The BFGS method, used to achieve self-consistency, involves solving for $\nu$ different parameters, such as $a_\Gamma$ and $a_\Delta$, which may not have the same dimensions.  The technique involves calculating a $\nu \times \nu$ Hessian matrix whose elements may have values that differ by many orders of magnitude.  It is numerically challenging to invert such a matrix.  Therefore, to make the problem tractable, we transform to dimensionless variables.  We have already identified the appropriate quantities for rescaling lengths in Eq.~(\ref{eq:dima}) and energies in Eq. (\ref{eq:dimE}).  The envelope functions and Gaussian form for the electron density may be simply expressed in these terms, as may the potentials and the Fermi constraint.

The Schr\"{o}dinger equations are solved by finite element methods.  The average dimensionless energy expectation values $\tilde{E}_\Gamma$ and $\tilde{E}_\Delta$ are used in the Fermi constraint.  These are readily computed by our finite element solver, via Eqs. (\ref{eq:epG}) and (\ref{eq:epD}) as 
\begin{eqnarray}
\tilde{E}_\Gamma &=& \tilde{\epsilon}_\Gamma - \frac{1}{2} \langle \tilde{V}_\Gamma \rangle_\Gamma ,\\
\tilde{E}_\Delta &=& \tilde{\epsilon}_ \Delta - \frac{1}{2} \langle \tilde{V}_ \Delta \rangle_ \Delta .
\end{eqnarray}
The parameters to be solved for in our 1D model are then $\{ \tilde{a}_\Gamma,\tilde{a}_\Delta,\beta_{\Gamma}\}$.  The valley splitting does not affect these solutions, and is therefore calculated \textit{post hoc}, as described in Sec. \ref{sec:VS}.

The final results are obtained numerically for the 1/4 ML case, giving $a_\Gamma=0.64$~nm and $a_\Delta =1.30$~nm.  The value of $\beta_{\Gamma}$ is slightly larger than the estimate $\beta_{\Gamma} \simeq 1/3$ obtained in Ref.~\onlinecite{qian05}, and larger than the estimated value of 0.310 from Ref. \onlinecite{oliver} (see Sec. \ref{sec:formalism}).

It should be noted that, with appropriate initial guesses as to the input parameters, the model converges in a matter of minutes on a standard laptop.  This is far more efficient than a typical \textit{ab initio} calculation, which often requires tens of hours' runtime across tens of cpus, and could potentially be used to model much larger systems, with device scales beyond the reach of more rigorous techniques.

\section{Exchange and Correlation}
\label{sec:XC}

The purpose of this Appendix is to detail when inclusion of correlation and exchange effects in the numerical model is necessary for accurate calculation.  It assumes the treatment given above, accounting for valley-splitting.  To characterize the effects of exchange and correlation, we perform our self-consistent calculations both with and without $V_{\text{X}}$ and $V_{\text{C}}$ in the Hamiltonian.

\begin{table}[t]
  \begin{center}
    \begin{tabular}{ccccc}
      \hline
      \hline
      Dopant density	& $a_{\Gamma}$	& $a_{\Gamma}$ 	& $a_{\Delta}$	& $a_{\Delta}$	\\
      (ML)						&	(nm)					& (nm, no XC)		&	(nm)					& (nm, no XC)		\\
      \hline
			1 							&	0.403			  	& 0.412  	    	& 0.825					&	0.876					\\
			2/3	      			& 0.461			  	& 0.472					& 0.943					&	1.007					\\
			1/2	      			& 0.508			  	& 0.521					& 1.036					&	1.112					\\
			1/3	      			& 0.581			  	& 0.598					& 1.183					&	1.280					\\
			1/4	      			& 0.639				  & 0.660					& 1.298					&	1.415					\\
			1/8	      			& 0.804				  & 0.837					& 1.625					&	1.804					\\
			1/16	      		& 1.009				  & 1.063					& 2.027					&	2.302					\\
			1/32	      		& 1.265				  & 1.350					& 2.521					&	2.948					\\
			1/64	      		& 1.580				  & 1.718					& 3.120					&	3.774					\\
			1/128	      		& 1.959				  & 2.185					& 3.854					&	4.852					\\
			1/256	      		& 2.420				  & 2.788					& 4.677					&	6.208					\\
			1/512	      		& 2.955				  & 3.547					& 5.601					&	7.993					\\
      \hline
      \hline
    \end{tabular}
  \end{center}
  \caption{Envelope widths ($a_{\Gamma,\Delta}$ values) with and without exchange and correlation.}
  \label{tab:avalues}
\end{table}

\begin{table}[b]
  \begin{center}
    \begin{tabular}{ccccccc}
      \hline
      \hline
      Dopant          & $\beta_{1}$   & $\beta_{1}$   & $\beta_{2}$   & $\beta_{2}$   & $\beta_{\Delta}$    & $\beta_{\Delta}$    \\
      density         &               & (no XC)       &               & (no XC)       &                     & (no XC)             \\
      (ML)            &               &               &               &               &                     &                     \\
      \hline
      1               & 0.169         & 0.169         & 0.146         & 0.147         & 0.685               & 0.684               \\
      2/3             & 0.174         & 0.174         & 0.157         & 0.158         & 0.670               & 0.669               \\
      1/2             & 0.178         & 0.178         & 0.164         & 0.165         & 0.658               & 0.657               \\
      1/3             & 0.186         & 0.186         & 0.175         & 0.175         & 0.640               & 0.639               \\
      1/4             & 0.192         & 0.192         & 0.182         & 0.183         & 0.626               & 0.625               \\
      1/8             & 0.210         & 0.209         & 0.202         & 0.202         & 0.588               & 0.590               \\
      1/16            & 0.231         & 0.228         & 0.225  	      & 0.223         & 0.544               & 0.549               \\
      1/32            & 0.255         & 0.249         & 0.251         & 0.245         & 0.494               & 0.506               \\
      1/64            & 0.284         & 0.272         & 0.280         & 0.269         & 0.437               & 0.459               \\
      1/128           & 0.317         & 0.295         & 0.314         & 0.293         & 0.370               & 0.412               \\
      1/256           & 0.357         & 0.319         & 0.354         & 0.317         & 0.289               & 0.365               \\
      1/512           & 0.408         & 0.341         & 0.406         & 0.339         & 0.186               & 0.320               \\
      \hline
      \hline
    \end{tabular}
  \end{center}
  \caption{Filling fraction ($\beta$ values) with and without exchange and correlation.}
  \label{tab:betas}
\end{table}

The primary effect of including exchange and correlation (XC), as discussed in the main text, is to slightly deepen the potential well and steepen it in the vicinity of the minimum.  This results in a contraction of the envelope functions, as can be seen in Table \ref{tab:avalues}.  The effect is more marked for the low-density cases, where the jellium approximation is in question and $\left(V_{\text{X}0}+V_{\text{C}0}\right)/V_{0}$ is closer to 1.  In the regime above 1/16 ML, the modified potential results in a change of less than 6\% in $a_{\Gamma}$, and less than 12\% in $a_{\Delta}$.  

\begin{table}[t]
  \begin{center}
    \begin{tabular}{ccccc}
      \hline
      \hline
      Dopant          & $\Gamma_{1}$-$\Gamma_{2}$  & $\Gamma_{1}$-$\Gamma_{2}$ & $\Gamma_{1}$-$\Delta$ & $\Gamma_{1}$-$\Delta$ \\
      density         & (meV)              				 & (meV)                     & (meV)                 & (meV)                 \\ 
      (ML)            &                    				 & (no XC)                   &                       & (no XC)               \\
      \hline
      1               & 183.2                      & 175.6                     & 595                   & 600                   \\
      2/3             & 88.3                       & 84.0                      & 446                   & 449                   \\
      1/2             & 54.1                       & 51.6                      & 363                   & 365                   \\
      1/3             & 28.8                       & 27.6                      & 271                   & 272                   \\
      1/4             & 19.0                       & 18.2                      & 220                   & 220                   \\
      1/8             & 7.3                        & 7.0                       & 132                   & 132                   \\
      1/16            & 2.8                        & 2.7                       & 79.2                  & 77.8                  \\
      1/32            & 1.1                        & 1.0                       & 47.1                  & 45.4                  \\
      1/64            & 0.44                       & 0.41                      & 27.9                  & 26.2                  \\
      1/128           & 0.18                       & 0.16                      & 16.4                  & 14.8                  \\
      1/256           & 0.07                       & 0.06                      & 9.7                   & 8.3                   \\
      1/512           & 0.03                       & 0.02                      & 5.8                   & 4.6                   \\
      \hline
      \hline
    \end{tabular}
  \end{center}
  \caption{Energy level splittings with and without exchange and correlation.}
  \label{tab:split}
\end{table}

\begin{table}[b]
  \begin{center}
    \begin{tabular}{ccc}
      \hline
      \hline
      Dopant density	& $E_{F}$				& $E_{F}$ 			\\
      (ML)						&	(meV)					& (meV, no XC)	\\
      \hline
			1 							&	474 			  	& 668   	    	\\
			2/3	      			& 251 			  	& 414 					\\
			1/2	      			& 151 			  	& 296 					\\
			1/3	      			& 63.1			  	& 185 					\\
			1/4	      			& 24.6				  & 132 					\\
			1/8	      			& -22.5				  & 57.7					\\
			1/16	      		& -35.7				  & 24.0					\\
			1/32	      		& -35.5				  & 8.97					\\
			1/64	      		& -30.6				  & 2.81					\\
			1/128	      		& -24.8				  & 0.32					\\
			1/256	      		& -19.3				  & -0.38					\\
			1/512	      		& -14.8				  & -0.57					\\
      \hline
      \hline
    \end{tabular}
  \end{center}
  \caption{Fermi energy values with and without exchange and correlation.  Note that the asymptotic value of the potential, $V\left(\infty\right)$, has been defined as the energy zero.}
  \label{tab:Fermi}
\end{table}


The filling fractions, $\beta_{1}$, $\beta_{2}$ and $\beta_{\Delta}$, are relatively insensitive to this perturbation.  For 1/4 ML doping density as described above, and indeed for densities above 1/16, the values are almost identical with and without exchange and correlation, although there is a marked difference at low densities.  Table \ref{tab:betas} details values of $\beta_{1}$, $\beta_{2}$ and $\beta_{\Delta}$.

Table \ref{tab:split} shows the effects of exchange and correlation, and density on the various energy level splittings.  As might be expected, the valley-splitting also shows little effect of the inclusion or exclusion of exchange and correlation for those doping densities where $V_{\text{XC}}$ is less significant.  For densities higher than 1/16 ML, the difference in the relative predicted splitting is less than 6\%.  Of course, if instead one is interested in the absolute difference in the predictions, then exchange and correlation lead to an increase in the splitting of more than 0.1 meV for all doping densities above 1/32 ML.

The $\Gamma_{1}$-$\Delta$ energy gap behaves similarly, with a relative difference of less than 2\% for systems denser than 1/16 ML.  As does the $\Gamma_{1}$-$\Gamma_{2}$ splitting, the energy gap increases with inclusion of exchange and correlation.  This is unsurprising, since the deeper potential corresponds to stronger confinement and hence wider spacings between energy levels.

Table \ref{tab:Fermi} displays the calculated Fermi energy with and without exchange and correlation.  It is of note that we observe Fermi energies greater than zero, corresponding to an unphysical ``overfilling'', over a much larger range of densities when ignoring exchange and correlation, with all densities above 1/128 ML (inclusive) having positive Fermi energies.  This effect is more pronounced for higher densities.

We note that, despite the Fermi energies changing by several meV for the 1/16-1/4 ML models, and being greater than zero when exchange and correlation are excluded, all other results are largely unaffected by the inclusion or exclusion of exchange and correlation in the calculation.  This is in line with the observations made in Ref. \onlinecite{qian05}, where the Fermi level was by far the measure most sensitive to exclusion of exchange and correlation.  We therefore observe that the overfilling appears to have little effect on these results - again in line with the discussion regarding the populations of higher bands in Ref. \onlinecite{qian05}.  We may then also consider it to have a similarly small effect on the higher densities, which also exhibit change due to the inclusion of exchange and correlation.

While to first order, the inclusion of exchange and correlation appears to have little effect on our main results, we would like to emphasize its contribution to second-order effects such as the Fermi energy.  We have noted that for several considered densities, including exchange and correlation reduces or even eliminates overfilling.  It can easily be imagined that, when connected in a physical device, significant overfilling would lead to breaking charge neutrality as the high-energy electrons are energetically free to vanish into the leads.  As charge neutrality is a central assumption in the derivation of our model, this is more important than perhaps it first appears.  We therefore recommend the inclusion of correlation and exchange in any EMT model of this type, especially if larger-scale device modeling is to be undertaken.

\end{document}